\documentclass[12pt]{article}

\usepackage{times}
\usepackage{fixltx2e}
\usepackage{graphicx}
\usepackage[table]{xcolor}
\usepackage{ragged2e}
\usepackage{multirow}
\usepackage{helvet}
\usepackage{booktabs}
\usepackage{colortbl}

\usepackage{blindtext}

\makeatletter
\newenvironment{figurehere}
{\def\@captype{figure}}
{}
\makeatletter

\usepackage{ccaption}

\newenvironment{sciabstract}{%
\begin{quote} \bf}
{\end{quote}}

\newcounter{lastnote}

\title{Deep Random Forest with Ferroelectric Analog Content Addressable Memory}

\author
{Xunzhao Yin$^{1}$, Franz M{\"u}ller$^{2}$, Ann Franchesca Laguna$^{3}$, Chao Li$^{1}$, 
Wenwen Ye$^{1}$, \\
Qingrong Huang$^{1}$, Qinming Zhang$^{1}$, Zhiguo Shi$^{1}$, Maximilian Lederer$^{2}$, \\
Nellie Laleni$^{2}$, Shan Deng$^{4}$, Zijian Zhao$^{4}$,  Michael Niemier$^{3}$, \\ 
Xiaobo Sharon Hu$^{3}$, Cheng Zhuo$^{1*}$, Thomas K{\"a}mpfe$^{2*}$, Kai Ni$^{4*}$
\\
\normalsize{$^{1}$Zhejiang University, Hangzhou, Zhejiang, China;}\\
\normalsize{$^{2}$Fraunhofer IPMS, Dresden, Germany;}\\
\normalsize{$^{3}$University of Notre Dame, Notre Dame, IN 46614, USA;}\\
\normalsize{$^{4}$Rochester Institute of Technology, Rochester, NY 14623, USA;}
\\
\normalsize{$^\ast$To whom correspondence should be addressed; E-mail:}\\ \normalsize{czhuo@zju.edu.cn, thomas.kaempfe@ipms.fraunhofer.de, kai.ni@rit.edu.}
}

\date{}

\begin{document} 

\maketitle 
\begin{sciabstract}
Deep random forest (DRF), which incorporates the core features of deep learning and random forest (RF), exhibits comparable classification accuracy, interpretability, and low memory and computational overhead when compared with deep neural networks (DNNs) in various information processing tasks for edge intelligence. 
However, the development of efficient hardware to accelerate DRF is lagging behind its DNN counterparts. 
The key for  hardware acceleration of DRF lies in efficiently realizing the branch-split operation at decision nodes when traversing a decision tree. 
In this work, we propose to implement DRF through simple associative searches realized with ferroelectric analog content addressable memory (ACAM). Utilizing only two ferroelectric field effect transistors (FeFETs), the ultra-compact ACAM cell can perform a branch-split operation with an energy-efficient associative search by storing the decision boundaries as the analog polarization states in an FeFET.
The DRF accelerator architecture and the corresponding mapping of the DRF model to the ACAM arrays are presented.
The functionality, characteristics, and  scalability of the FeFET ACAM based DRF and its robusteness against FeFET device non-idealities are validated both in experiments and simulations.  
Evaluation results show that the FeFET ACAM DRF accelerator exhibits $\sim$10$^6$x/16x and $\sim$10$^6$x/2.5x improvements in terms of energy and latency when compared with other deep random forest hardware implementations on the state-of-the-art CPU/ReRAM, respectively.

\end{sciabstract}
\section*{Introduction}
\label{sec:introduction}
Edge intelligence in the era of Internet of Things (IoT) requires that  raw data is analyzed locally instead of being transmitted back to the cloud for processing \cite{keshavarzi2019edge,keshavarzi2020ferroelectronics, zhou2019edge}. 
Such edge intelligence can best be achieved by deploying an artificial intelligence (AI) hardware engine designed for IoT devices. 
Deep neural networks (DNNs) are highly effective in processing visual and speech data for various applications with high accuracy. However, DNN models face several fundamental challenges, and are not readily deployable in the IoT. First, modern DNN models require large memories to store learned weights (commonly $>$1GB) \cite{xu2018scaling}, well-beyond the capacity of an embedded, on-chip memory in edge devices. External memories are therefore needed to store the entire DNN model. 
The requisite data transfers between on/off-chip memory leads to significant energy and latency overheads which in turn limit the network complexities that may be deployed in edge devices. 
Second, to achieve high accuracy, DNNs require a significant amount of labeled training data. Data collection and preparation is expensive and time-consuming for many tasks -- especially for edge devices, considering their diverse functionalities and applications \cite{wang2020convergence,xu2020edge,feng2018edge}. 
Third, the "black box" nature and large parameter space of a DNN makes it challenging to analyze and understand how DNNs make their decisions.  In certain domains, such as medicine, health care, and finance, the interpretability of a model is critical in establishing trust and developing solutions to other related problems \cite{doshi2017towards, chakraborty2017interpretability, vellido2019importance, arrieta2020explainable}. 
In light of these challenges, deep random forests (DRF), a recently proposed interpretable and memory-efficient AI model \cite{zhou2019deep}, are considered to be an excellent alternative to DNNs in realizing light-weight AI engines for edge intelligence.

At a high level, DRF incorporates the core features of deep learning models, i.e., layer-by-layer processing, in-model feature transformation, and sufficient model complexity \cite{zhou2019deep}, as shown in Fig. \ref{fig:overview}(a). DRF follows a cascaded structure where each layer in a DRF receives feature information extracted from the preceding level. Each layer is an ensemble of random forests (i.e., an ensemble of weak decision tree based classifiers). Each forest models the class distribution of the datasets either through majority voting or averaging the predictions of decision trees in the same random forest. Those outputs from the forests in the same layer are concatenated together and forwarded to the next layer for further processing \cite{zhou2019deep}. Equipped with these deep model features, DRF achieves comparable or better accuracy with DNNs in processing low-resource dataset \cite{zhou2019deep}). In addition, by inheriting the interpretability and low energy and memory requirements of the random forest \cite{fernandez2014we}, DRF represents a competitive solution for edge intelligence to handle information processing tasks with requirements that DNNs might struggle to satisfy (e.g., limited resources or interpretability). Unlike DNNs,  hardware acceleration of DRF has not been well explored. Our work addresses this gap by introducing an energy-efficient and high performance hardware for accelerating DRF. 

The key challenge in accelerating DRF is to implement the decision trees, the core component of DRF, as shown in Fig. \ref{fig:overview}(b). It perform comparisons at each non-leaf node, and depending on the comparison results, the node is split into different branches. It has been proposed that ACAM can be used to perform the branch-split operation in a decision tree \cite{pedretti2021tree}, which opens up the possibility of accelerating DRF with ACAM arrays.
As a type of associative memories, CAMs have gained popularity in data-centric computing due to their massively parallel pattern-matching capability \cite{pagiamtzis2006content, karam2015emerging}. 
They can identify the stored entries matching the search query in parallel in the exact or approximate matching mode. In the exact matching mode, only the items that exactly match the input query are identified \cite{karam2015emerging}, 
while in the approximate matching mode, the Hamming distance (HD) between the query and stored entries are returned by sensing the match-line (ML) current. The approximate matching function has been applied to accelerate various machine learning applications \cite{imani2017exploring,ni2019ferroelectric}. All the developments above have only considered digital CAMs, where binary information is stored and searched. 
However, it is also possible to leverage the analog states of nonvolatile memories for multi-bit or ACAMs \cite{yin2020fecam,li2020analog, li2020scalable}. Multi-bit information can be stored in the CAM, and an analog or multi-bit query can be searched across the CAM array for pattern matching, thus greatly improving the information density and expanding the CAM functionality \cite{yin2020fecam,li2020analog, li2020scalable, kazemi2020memory}. In this work, we demonstrate ferroelectric ACAMs and leverage their unique properties to accelerate DRF.

\newpage
\begin{figurehere}
 \begin{center}
  \includegraphics[width=1\textwidth]{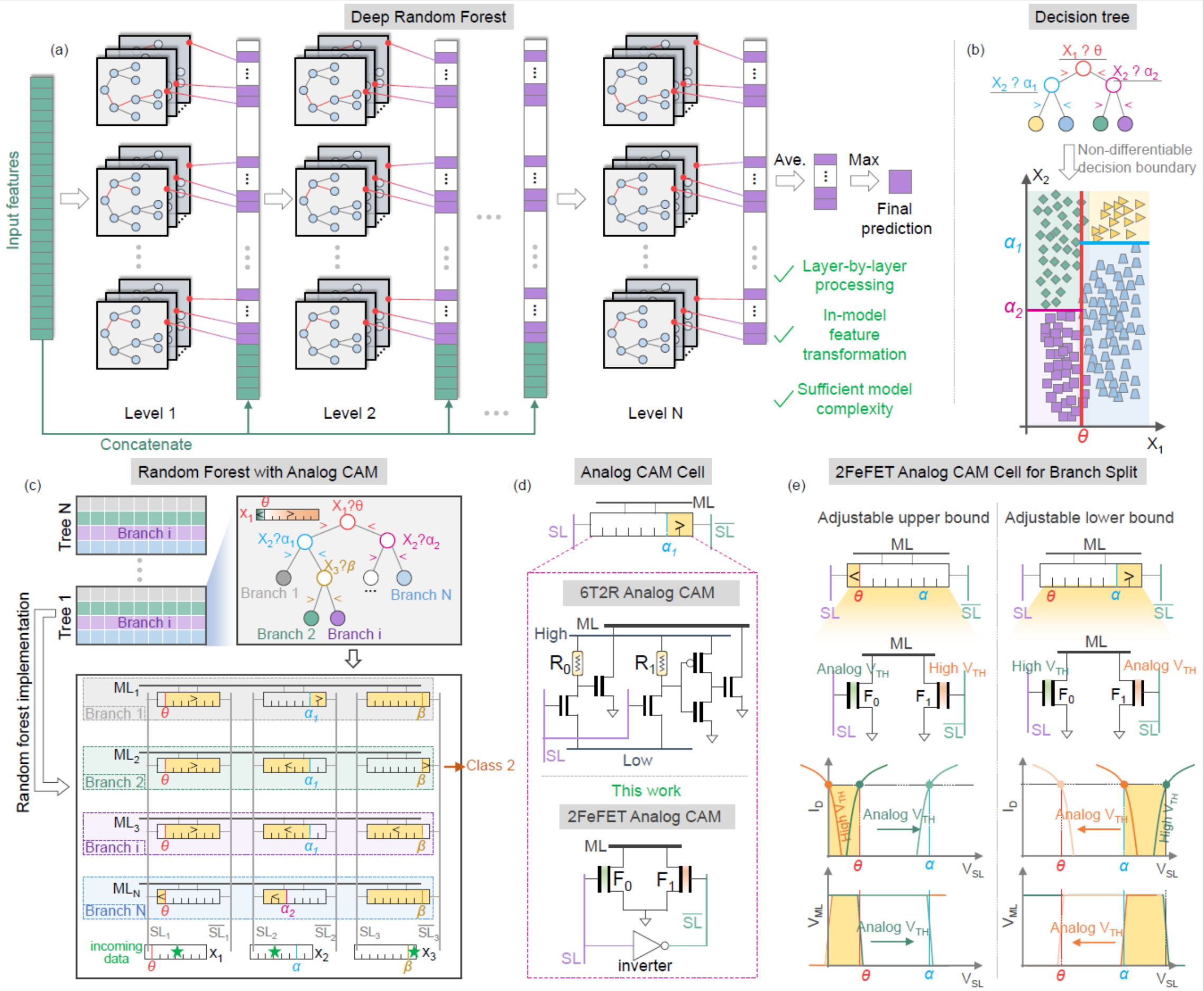}
  \caption{\textit{Overview of implementing DRF with ferroelectric ACAM. (a) DRF is a deep model built by cascading random forests, forming a layer-by-layer structure. The output of each layer concatenates a portion of the input features, allowing in-model feature transformation. The resulting DRF model can achieve good performance. (b) Each decision tree in a random forest forms a non-differentiable decision boundary by making a branch split at each non-leaf node based on the input features. (c) The random forest can be mapped onto an ACAM array. An ACAM cell with adjustable matching bounds (i.e., upper or lower matching bound) can efficiently realize the branch-split operation in a decision tree; as such an ACAM word can realize a branch from the root node to the leaf node in a decision tree. (d) The existing demonstrated ACAM cells based on the multi-bit embedded nonvolatile memories. Compared with its 6T2R ReRAM ACAM counterpart, 2FeFET based ACAM is compact and universal by simultaneously serving as a digital and analog CAM. (e) The working principle of 2FeFET ACAM cell with adjustable upper/lower matching bound to realize the branch split in a decision tree.}}
  \label{fig:overview}
 \end{center}
\end{figurehere}

In an ACAM cell, a matching range, defined by the upper and lower bounds of the search line (SL) voltage, can be dynamically adjusted by configuring the memory device states \cite{yin2020fecam}. We observe that by fixing the upper/lower bound of the matching range to the maximum/minimum voltage allowed on the SL and leaving the corresponding lower/upper bound adjustable, the respective greater-than (i.e., $>$)/less-than (i.e., $<$) branch-split operations in a decision tree can be efficiently implemented in an ACAM cell through a simple search operation, as shown in Fig. \ref{fig:overview}(c). An ACAM word, composed of a row of CAM cells, can be used to implement a branch from the root node to a leaf node in a decision tree, while an ACAM array represents an entire decision tree. In this way, the decision space partitioned by the decision tree can be mapped into the matching space of an ACAM array.  
As a result, the inference operation of a decision tree can be realized through a simple parallel search operation in an ACAM array. The identified matched entries indicate the prediction results (i.e., the matching branches). By cascading multiple ACAM arrays together, the DRF can be realized. The effects of the limited precision of ACAM cells in defining the decision boundaries and the device-to-device variation of ACAMs 
 are explored in the system benchmarking section.

Developing ACAM arrays for DRF requires that the ACAMs be compact, fast and energy efficient.
In our previous work \cite{yin2020fecam}, we have proposed a universal ferroelectric CAM design through SPICE simulations, in which a CAM cell composed of two ferroelectric FETs (FeFETs) can simultaneously serve as a digital and analog CAM cell. Notably, the 2FeFET CAM is the most compact cell to date, compared with SRAM based CAM cells typically composed of 16 transistors, spin-transfer-torque magnetic random access memory (STT-MRAM) based CAM cells built using 10-15 transistors and 2-4 magnetic tunnel junctions (MTJ), and a resistive memory (i.e., Resistive random access memory (ReRAM) and phase change memory (PCM)) based CAM cell constructed with 2 transistors and 2 resistive memory devices \cite{ni2019ferroelectric}. Additionally, CAM based on FeFET is especially energy efficient. Unlike a volatile SRAM CAM which consumes a significant leakage power, FeFET CAM is nonvolatile, thus avoiding the energy consumption due to leakage current.  Moreover, unlike other NVMs where switching is typically driven by a large conduction current, ferroelectric switching can be induced with an applied electric field without consuming conduction current, thus exhibiting superior energy efficiency. Write energy down to 1fJ/bit is achievable in a single FeFET \cite{keshavarzi2020ferroelectronics, schroeder2020nonvolatile}. 
Finally, ferroelectric CAM exhibits superior performance owing to its intrinsic transistor structure and a large \textit{I}\textsubscript{ON}/\textit{I}\textsubscript{OFF} ratio (e.g., $\sim$10$^4$), significantly outperforming the two-terminal resistive memories which typically show an \textit{I}\textsubscript{ON}/\textit{I}\textsubscript{OFF} ratio of $\sim$100. 
These characteristics enables 2FeFET CAM to simultaneously serve as both a digital and an analog CAM, creating a versatile hardware platform for various applications. 

In this work, we demonstrate the 2FeFET based ACAM for the implementation of a DRF. 
There have been reports of utilizing other NVM devices to implement an ACAM cell, such as the firstly proposed ReRAM ACAM \cite{li2020analog} (Fig. \ref{fig:overview}(d)). However, due to its limited \textit{I}\textsubscript{ON}/\textit{I}\textsubscript{OFF} ratio, additional transistors are added into the digital CAM cell core (e.g., the 2T2R CAM cell) to support the analog/multi-bit search functionality, making it a 6T2R structure \cite{li2020analog}, larger than the 2FeFET ACAM design. The operating principles of the proposed 2FeFET ACAM cell for implementing the branch-split operation in a decision tree are illustrated in Fig. \ref{fig:overview}(e). To implement a less-than branch (Fig. \ref{fig:overview}(e)(left)), the FeFET \textit{F}\textsubscript{1}, connected with  $\overline{\mbox{SL}}$, is set to the high-\textit{V}\textsubscript{TH} state such that it remains in the cut-off state over the entire SL search range, thus forming a fixed lower bound. 
Adjusting the \textit{V}\textsubscript{TH} state of the FeFET \textit{F}\textsubscript{0} associated with the SL tunes the upper bound of the matching range. 
When the SL search voltage \textit{V}\textsubscript{SL} falls within the yellow region (where both the FeFETs turn off and the ML discharges slowly), the ML voltage, \textit{V}\textsubscript{ML}, remains high throughout the sensing phase of a voltage sense amplifier. 
When the \textit{V}\textsubscript{TH} of \textit{F}\textsubscript{0} increases, the resulting upper bound of the matching range also increases. 
As a result, the less-than branch with different thresholds can be mapped to the 2FeFET ACAM cell with an adjustable upper bound.
By symmetry, the greater-than branch can also be achieved by setting \textit{F}\textsubscript{0} in high-\textit{V}\textsubscript{TH} state, forming a fixed upper bound and adjusting the \textit{V}\textsubscript{TH} of \textit{F}\textsubscript{1} to set the lower bound of the matching range. 
For the cases where not all the branches are of the same length, such as branch 1 and branch 2 in Fig. \ref{fig:overview}(c), or of the same set of features for branch split, the 'don't care' functionality of ACAM is leveraged. 
When a branch-split operation occurs over an input feature that is not included in the other branches, the ACAM cells mapping the missing features in those branches are set to the 'don't care' state so that they contribute negligible leakage current through the ML, without impacting the \textit{V}\textsubscript{ML}. 
The 'don't care' functionality can be realized by simply setting both FeFETs of the ACAM cell to the high-\textit{V}\textsubscript{TH} state.

In the following sections, we first describe the experimental demonstration of the 2FeFET ACAM cell and verify the branch-split operation for decision trees. 
We also demonstrate the capability of an ACAM word in realizing a branch from the root node to a leaf node in a decision tree through a simple search operation. 
This capability is utilized to realize a DRF, which exhibits good performance and superior energy efficiency
. In addition, we present the evaluation of the impact of FeFET non-idealities, such as variation and limited precision, on the DRF performance to demonstrate the robustness of FeFET ACAM DRF. The main contributions of the paper are: i) proposing a DRF accelerator leveraging the ferroelectric ACAM arrays for edge intelligence; ii) first experimental demonstration of a ultra-compact, energy-efficient, and universal 2FeFET digital and analog CAM cell; iii) demonstrating the capability of an ACAM array in mapping  a decision tree to the matching space of that ACAM array; iv) evaluating the impact of limited FeFET bit precision on the accuracy of DRF and proposing a precision extension method using low-precision devices; v) demonstrating the significant robustness of DRF against device-to-device variation.

\section*{2FeFET Analog CAM Demonstration}
\label{sec:cam}

In this section, we first discuss the experimental validation of the ACAM cell operation. 
We have constructed the proposed ACAM cell with the industrial 28 nm high-k metal gate (HKMG) FeFET technology (shown in Fig. \ref{fig:singlecell_exp}(a) and (b)). The device features an 8 nm thick doped HfO$_2$ ferroelectric thin film as the gate dielectric, capped with a TiN and polysilicon layer. 
A thin SiO$_2$ interlayer ($\sim$1 nm) is also present between the ferroelectric and the silicon substrate. Fig. \ref{fig:singlecell_exp}(b) shows the schematic cross-section of the device. 
Detailed process information can be found in \cite{trentzsch201628nm}. 
The local crystallographic phase has been characterized in the ferroelectric HfO$_2$ films by
transmission-electron back-scattering
diffraction (EBSD) \cite{lederer2019local}, as shown in Fig. \ref{fig:singlecell_exp}(c). Dendritic grains consisting of the ferroelectric orthorhombic phase are observed and only a small portion of the film grains are in the monoclinic dielectric phase, suggesting a good control over the ferroelectric phase through the high temperature stressed annealing. 
From the in-plane inverse pole figure map (Fig. \ref{fig:singlecell_exp}(d)) a large variety of crystallographic orientations can be deduced. 
As a consequence, the polarization axis in each grain will be located at slightly different angles. 
Moreover, as gradients can be observed inside these grains and especially the dendrites, high degrees of intra-granular misorientation are expected \cite{lederer2021impact}. 
Consequently, these dendrites are likely to switch at slightly different electric fields and are therefore reducing the effective grain size of the film. A broad distribution of polarization orientations as well as small switchable regions, as present in this film, allows for analog-like multi-state operation in the ferroelectric HfO\textsubscript{2} layer.

The FeFET \textit{I}\textsubscript{D}-\textit{V}\textsubscript{G} characteristics for the low-\textit{V}\textsubscript{TH} and high-\textit{V}\textsubscript{TH} states after $\pm$4 V, 1 $\mu$s write pulses are shown in Fig. \ref{fig:singlecell_exp}(e). Device variation is characterized by measuring 60 different devices. The results show a large memory window of $\sim$1.2 V and a large sensing margin (i.e., \textit{I}\textsubscript{ON}/\textit{I}\textsubscript{OFF}) separating the two \textit{V}\textsubscript{TH} states even when considering the device variation. 
The switching dynamics of the tested FeFET are shown in Fig. \ref{fig:singlecell_exp}(f), where the required pulse width to obtain a memory window of 1.2 V as a function of write pulse amplitude is presented. The required switching time can be well described by the expression derived from domain nucleation theory \cite{mulaosmanovic2017switching,mulaosmanovic2020investigation}
\[PW=\tau_o e^{\frac{\alpha}{(V_w-V_{off})^2}}\]
where \(\alpha\) is a fitting parameter related with the polarization switching barrier, \(\tau_o\) is the switching time at an infinitely large applied pulse amplitude, and \(V_{off}\) is the offset voltage, an indication of the local domain environment. 
With the increase of write pulse amplitude, FeFET switching speed can be further reduced to below 10 ns \cite{bae2020sub}, suggesting the great promise for high speed and energy-efficient ferroelectric memory. 

\begin{figurehere}
 \begin{center}
  \includegraphics[width=1\textwidth]{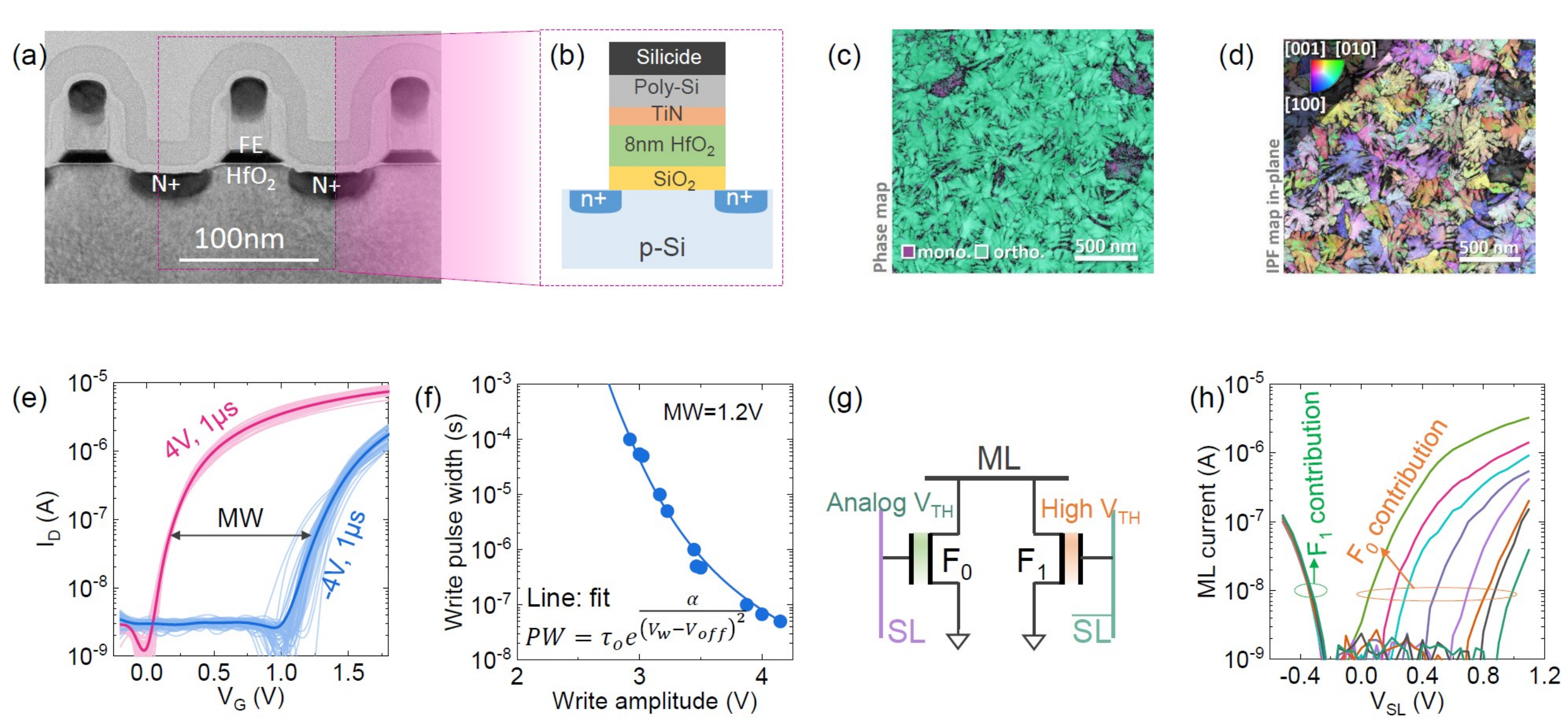}
  \caption{\textit{Experimental demonstration of a ferroelectric ACAM cell. (a) The cross-sectional TEM image of the FeFET device and (b) its schematic cross-section. It features an 8 nm thick doped HfO$_2$ ferroelectric film. (c) The phase analysis through transmission-EBSD confirms that the poly-crystalline HfO$_2$ film consists mostly of the orthorhombic ferroelectric phase. Inverse pole figure maps (d) reveal intra-granular misorientation, especially in the dendrites. (e) The experimentally measured \textit{I}\textsubscript{D}-\textit{V}\textsubscript{G} characteristics for low-\textit{V}\textsubscript{TH} and high-\textit{V}\textsubscript{TH} states after $\pm4$ V, 1 $\mu$s write pulses. 60 different devices are measured, suggesting excellent device variation control in the FeFET. (f) The representative switching dynamics in the FeFET. To obtain a given memory window (e.g. 1.2 V in this case), the required switching time as a function of applied pulse amplitude can be well-fitted with the nucleation limited switching model. (g) The CAM cell configuration used in the experimental validation, where F$_1$ is set to be highest \textit{V}\textsubscript{TH} state and the F$_0$ is adjusted. (h) Measured ML current as a function of the search line voltage, \textit{V}\textsubscript{SL}. Since F$_1$ is fixed to be highest \textit{V}\textsubscript{TH}, it contributes negligible current. When the  \textit{V}\textsubscript{TH} of F$_0$ is varied, the threshold of the matching range is shifted, thus demonstrating successful single cell operation.}}
  \label{fig:singlecell_exp}
 \end{center}
\end{figurehere}

Leveraging the partial polarization switching in the multi-domain FeFET, multiple \textit{V}\textsubscript{TH} states have been demonstrated and utilized for multi-level cell memories and synaptic weight cells for the acceleration of matrix-vector multiplication \cite{jerry2017ferroelectric, sun2018exploiting, halter2020back}. 
In this work, we harness the intermediate \textit{V}\textsubscript{TH} states to realize the branch-split operation with adjustable thresholds for the non-leaf nodes in a decision tree for DRF. 
Fig. \ref{fig:supple_fefet_states} shows experimentally measured \textit{I}\textsubscript{D}-\textit{V}\textsubscript{G} characteristics for 4 \textit{V}\textsubscript{TH} levels in a FeFET, which are set by applying different pulse amplitudes. 
The extracted \textit{V}\textsubscript{TH} distribution for 4 levels and 8 levels are shown in Fig. \ref{fig:supple_fefet_states}(c) and Fig. \ref{fig:supple_fefet_states}(d), respectively. 
With negligible overlaps between the neighboring levels, it is feasible to store multiple states into a FeFET, thus enabling the ACAM application proposed in this work. 
As shown in Fig. \ref{fig:singlecell_exp}(g), to verify the single ferroelectric ACAM cell operation, the FeFET associated with $\overline{\mbox{SL}}$ (\textit{F}\textsubscript{1}) is set to the high-\textit{V}\textsubscript{TH} state (\textit{V}\textsubscript{TH}=1.1V) and the FeFET associated with SL (\textit{F}\textsubscript{0}) is configured to different \textit{V}\textsubscript{TH} states.
The ML current is then measured with a sweeping SL voltage, \textit{V}\textsubscript{SL}. As a result, the matching range of \textit{V}\textsubscript{SL} where the ML current is low can be identified. 
Such \textit{V}\textsubscript{TH} configurations defines a matching range with varying upper bounds over the \textit{V}\textsubscript{SL}, thus implementing  a less-than branch-split operation with varying decision boundaries. 
Due to the symmetry of the ACAM cell, the greater-than branch split is realized by simply swapping the \textit{V}\textsubscript{TH} settings of the two FeFETs. Fig. \ref{fig:singlecell_exp}(h) shows the measurement results corresponding to Fig. \ref{fig:singlecell_exp}(g). With the high-\textit{V}\textsubscript{TH} state of \textit{F}\textsubscript{1}, this FeFET is cut-off in the entire voltage range (i.e., 0V to 1V), and is only turned on at negative \textit{V}\textsubscript{SL}. 
By setting \textit{V}\textsubscript{TH} of \textit{F}\textsubscript{0} to 8 different states, the upper bounds for the matching range is defined accordingly. As such, this verifies the successful operation of the ferroelectric ACAM cell.

To exploit a ferroelectric ACAM word for the mapping of an entire branch from the root node to a leaf node of a decision tree, we further validate the capability of an ACAM word to define a matching subspace in the high dimensional feature space spanned by the \textit{V}\textsubscript{SL} inputs of all the ACAM cells. 
Fig. \ref{fig:array_exp}(a) illustrates the experimental validation of the ferroelectric ACAM word. Fig. \ref{fig:array_exp}(b) shows the compact layout for an ACAM word.
Without loss of generality and for better illustration, a 1$\times$2 ACAM word consisting of two ACAM cells is demonstrated, which can define a matching subspace in the whole feature space spanned by the \textit{V}\textsubscript{SL1} and \textit{V}\textsubscript{SL2}. 
For the experimental demonstration, similar to the single cell case, the \textit{F}\textsubscript{1} transistor in both cells is set to the high-\textit{V}\textsubscript{TH} state while the \textit{V}\textsubscript{TH} state of \textit{F}\textsubscript{0} is varied among 4 different levels from 0 V to 1.1 V. 
Since in the ACAM array, each cell is independent from each other. As such, the \textit{V}\textsubscript{TH} of \textit{F}\textsubscript{0} defines a   \textit{V}\textsubscript{SL} plane, below which the cell contributes negligible current, indicating a match. 
When multiple cells are connected in parallel on the same ML, each cell defines one such \textit{V}\textsubscript{SL} plane, and the intersection of the space bounded by those planes defines the matching subspace of the ACAM word, namely the search input space that satisfies all the split conditions along a branch of a decision tree. Fig. \ref{fig:array_exp}(c) illustrates the ML current as a function of \textit{V}\textsubscript{SL1} and \textit{V}\textsubscript{SL2} when the \textit{V}\textsubscript{TH} of \textit{F}\textsubscript{0} in both cell 1 and cell 2 is set to one of the 4 different levels  (a total of 4x4 configurations). 
The 3D colormap surface of the ML current and its projection on the \textit{V}\textsubscript{SL1} and \textit{V}\textsubscript{SL2} plane are presented. 
It clearly indicates that the low current region on each dimension (e.g., $\leq$10$^{-7}$A in this work) follows the \textit{V}\textsubscript{TH} states of the \textit{F}\textsubscript{0} transistor in the corresponding cell. This successfully demonstrates the independence among the ACAM cells. Thus, the configured cell threshold sets the boundary of the matching subspace on the dimension of the corresponding cell. 
\begin{figurehere}
 \begin{center}
  \includegraphics[width=1\textwidth]{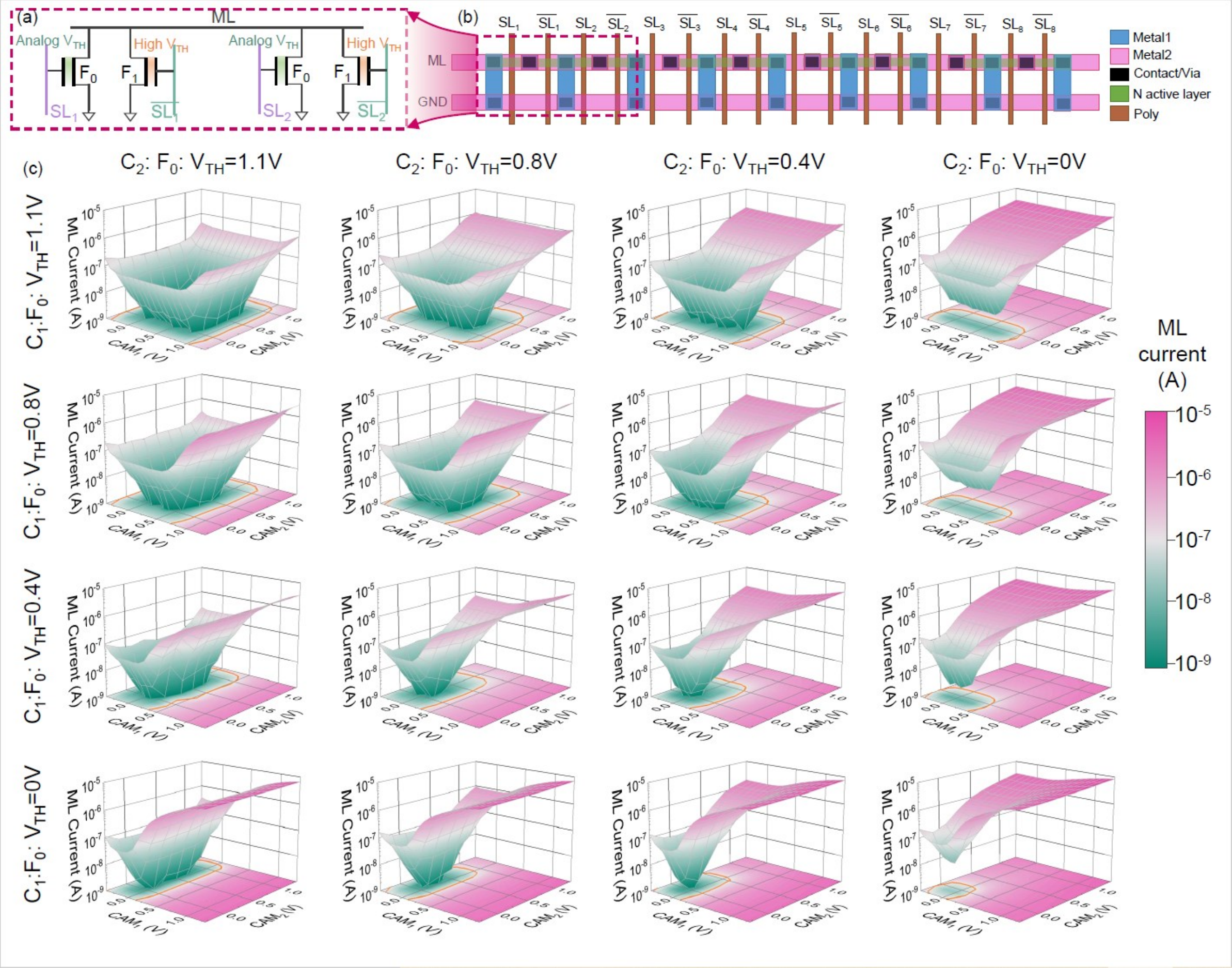}
  \caption{\textit{Experimental demonstration of ferroelectric ACAM array. (a) The configuration of FeFETs in the 1x2 ACAM word. F1 transistors in both cells are set to the high-\textit{V}\textsubscript{TH} state and F$_0$ transistors in both cells are configured to different \textit{V}\textsubscript{TH} states which set the threshold for the branch-split operation. (b) The compact layout of a 1x8 2FeFET ACAM word. (c) The experimental results show that the low ML current region (i.e., matched condition) can be configured in different locations in the \textit{V}\textsubscript{SL} space. Orange lines in each figure correspond to a match line current of $10^{-7}$ A. It successfully demonstrates the capability of ferroelectric ACAM word in configuring the matching subspace in the overall \textit{V}\textsubscript{SL} space.}}
  \label{fig:array_exp}
 \end{center}
\end{figurehere}
\newpage

An ACAM word with a larger size of 1$\times$16 has also been tested. As the matching subspace of the word lies in the 16-dimensional space and cannot be visualized, for ease of illustration, we consider a configuration where 15 cells are grouped together by storing the same state and are searched with the same information. 
The \textit{F}\textsubscript{1} transistors in all ACAM cells are in the high-\textit{V}\textsubscript{TH} state, enabling all cells to perform the lower-than branch split operation. The \textit{F}\textsubscript{0} transistors in the grouped 15 cells are set to the same intermediate state. The remaining single cell is adjusted among the four different \textit{V}\textsubscript{TH} states. 
After configuring the cells, the \textit{V}\textsubscript{SL} of the single cell and that of the grouped cells are swept from -0.3 V to 1.2 V in steps of 0.1 V. 
Fig. \ref{fig:cellarray_1x16_s4}, \ref{fig:cellarray_1x16_s3}, \ref{fig:cellarray_1x16_s2}, and \ref{fig:cellarray_1x16_s1} show the measured ML current when the \textit{F}\textsubscript{0} transistors of the grouped 15 cells are set to  \textit{V}\textsubscript{TH}=1.1 V, 0.8 V, 0.4 V, and 0 V, respectively. It clearly shows that on the dimension of each \textit{V}\textsubscript{SL}, the low ML current matching range closely follows the \textit{V}\textsubscript{TH} of the corresponding cell. This indicates that the boundary of the matching subspace on one \textit{V}\textsubscript{SL} dimension in the high-dimensional space is set by the decision boundary of that particular ACAM cell. 
This verifies the basic operation principles of the proposed ACAM array in realizing the branch-split operation of a decision tree in a DRF.  

To employ an ACAM array, voltage domain sensing is typically adopted for its simplicity, where the sense amplifier output voltage remains high when the search information matches the stored ACAM word; otherwise, the ML voltage discharges to ground. Such functionality has also been validated in SPICE simulations using a  calibrated FeFET compact model \cite{ni2018circuit} as shown in Fig. \ref{fig:spicesim}. In this work, a single two-stage buffer circuit is adopted for voltage domain sensing, where the output is binary, as shown in Fig. \ref{fig:sim_system_setup}. The output is close to \textit{V}\textsubscript{DD} when a low current flows through the ML (i.e., match case) and at ground when a mismatch happens. 
Fig. \ref{fig:spicesim}(a) shows the simulated ML current of a single cell configured to perform the less-than branch-split operation. The simulated ML current shows a similar trend as the experimental results shown in Fig. \ref{fig:singlecell_exp}(h). 
With this ML current dependence on \textit{V}\textsubscript{SL},  voltage domain sensing can be performed with the sense amplifier (SA) shown in Fig. \ref{fig:sim_system_setup}(a). The simulated output transient waveforms at different search voltages are shown in Fig. \ref{fig:single_cell_sense}. For \textit{V}\textsubscript{SL} in the matching subspace, the ML current is low; thus, ML voltage remains high. Otherwise, the ML voltage discharges to ground at a fast rate. 
At a certain sense time (e.g., in this work 10 ns is chosen), the SA output voltage varies as a function of \textit{V}\textsubscript{SL}, and multiple voltage thresholds for the branch-split operation can be defined depending on the stored \textit{V}\textsubscript{TH} in the cell, as shown in Fig. \ref{fig:spicesim}(b). Therefore, whether the input query matches with the defined branch condition can be determined by the output of the SA.   

\begin{figurehere}
 \begin{center}
  \includegraphics[width=1\textwidth]{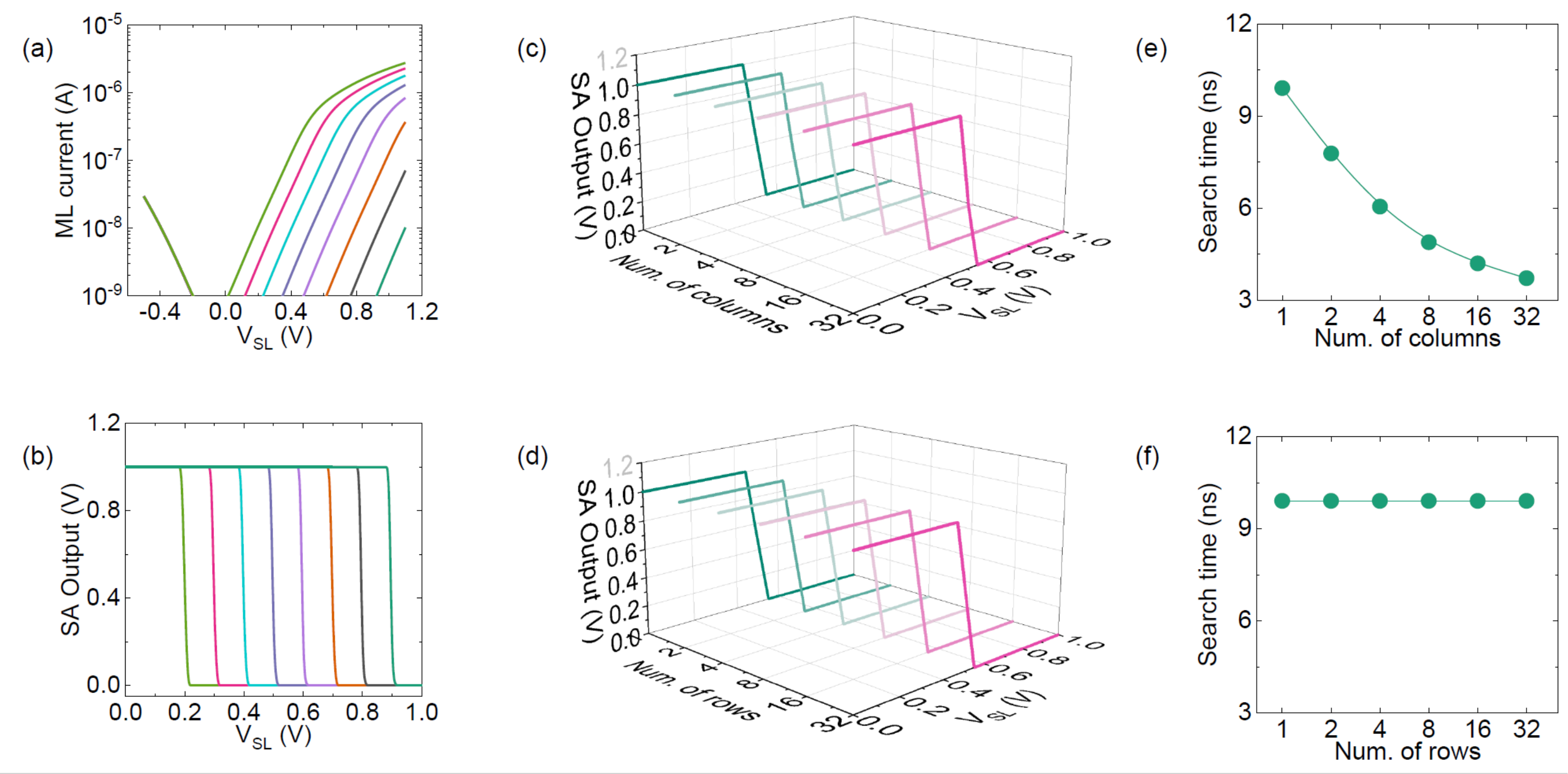}
  \caption{\textit{SPICE simulation of the ACAM cell and array. (a) The ML current of a single ACAM cell under different \textit{V}\textsubscript{SL} when the ACAM cell is configured to perform the less-than branch-split operation. (b) The output voltage of the two-stage buffer sense amplifier at the search time of 10 ns for the less-than branch-split operation. (c) and (d) The transfer characteristics of the sense amplifier output over the input voltage as a function of the number of columns and rows in the ACAM array, respectively. (e) and (f) The corresponding search time has to be adjusted for different number of columns (e), although the search time is almost the same for different number of rows (f).}}
  \label{fig:spicesim}
 \end{center}
\end{figurehere}

The operations of the ACAM array are also simulated.
Similar to the experiment shown in Fig. \ref{fig:array_exp}, the ML current of a 1$\times$2 ACAM array is simulated by sweeping  \textit{V}\textsubscript{SL1} and \textit{V}\textsubscript{SL2} of the cells. By setting  \textit{V}\textsubscript{TH} of  \textit{F}\textsubscript{0} in both cells in one of 4x4 configurations, different match subspaces can be realized in the space spanned by the \textit{V}\textsubscript{SL1} and \textit{V}\textsubscript{SL2} (as shown in Fig. \ref{fig:sim_1x2_array_current}, following the same behavior as the experiment shown in Fig. \ref{fig:array_exp}). 
Voltage domain sensing of the ACAM array is also implemented using the same setup as the single cell as shown in Fig. \ref{fig:sim_system_setup}(b). The impact of the array size (i.e., rows and columns of the ACAM array), on the voltage sensing of the ACAM array has been simulated, as shown in Fig. \ref{fig:spicesim}(c-f). 
A worst-case scenario is considered, where only one cell in the array is swept while all the other cells are searched with a \textit{V}\textsubscript{SL} close to the decision boundary, which makes it challenging to sense.
The impact of the number of columns (i.e., the number of ACAM cells connected to the same match line) on the sensing of the ACAM array is studied. 
As the number of cells per word increases,  the leakage current contributed by the cells searched close to the boundary becomes larger, resulting in an increased discharge rate of the match line. 
Therefore, as shown in Fig. \ref{fig:sim_array_col}, when the column size increases from 1 to 32, the search time needs to be adjusted accordingly.
Fig. \ref{fig:spicesim}(c) shows the output voltage as a function of  \textit{V}\textsubscript{SL1} for ACAM arrays with different number of columns sensed at the search times shown in Fig. \ref{fig:spicesim}(e). 
It can be seen that the decision boundary can be maintained across various sizes of arrays. 
Since the array size is pre-determined, the adjustment of sense time is straightforward. Fig. \ref{fig:spicesim}(d) and (f) show that the impact of the number of rows, i.e., number of ACAM words or independent match lines, on the array sensing is negligible, as each ML sensing is independent. Therefore, the decision boundary can be maintained for scaled array sizes.

\section*{Application Evaluation and Benchmarking} \label{sec:benchmarking}

Leveraging the validated FeFET ACAM array, the performance of DRF can be evaluated. The mapping of a DRF involves multiple ACAM arrays.
As demonstrated in Fig. \ref{fig:overview}, DRF is a machine learning framework that follows a layer-by-layer structure using cascaded random forests. Each layer is composed of multiple random forests, which output a probability for each class. A random forest uses an ensemble of decision trees to determine the probability of each target class for a given test example. 
Each decision tree can be mapped to an ACAM array as shown in Fig. \ref{fig:system_benchmarking}(a). Each cell represents a non-leaf node that performs the branch-split operation over a specific feature. Each row of the ACAM implements a branch from the root node to a leaf node. Hence, the number of rows corresponds to the number of leaf nodes (i.e., number of branches). 
The number of columns in an ACAM array corresponds to the number of features. Multiple ACAM arrays can be cascaded horizontally as shown in Fig. \ref{fig:system_benchmarking}(a) to hold all the features of a decision tree. As each cell in an ACAM word is independent of each other, a large ACAM word can be decomposed into multiple small ACAM words such that searching for a matching large ACAM word is equivalent to searching for the matching words in all ACAM subarrays simultaneously. 
Each ACAM array corresponding to a decision tree votes for a given class, and using a vote counter, the random forest outputs a vote vector which represents the number of votes for each class. 
The vote vectors of the random forests are then concatenated and passed to the next layer of the DRF.

DRF have been used in a variety of applications such as facial age estimation \cite{guehairia2020feature}, malware detection \cite{DRF_Malware} and classification of hyperspectral images \cite{cao2019densely}. Here we use two representative datasets for benchmarking to evaluate the accuracy of the DRF model. One is a image dataset, MNIST \cite{lecun1998gradient}, and the other is the time-series dataset, sEMG, used for hand movement recognition \cite{sapsanis2013improving}. The sEMG dataset consists of 1,800 records, where each one belongs to one of six hand
movements, i.e., spherical, tip, palmar, lateral, cylindrical, and hook.
Fig. \ref{fig:system_benchmarking}(b) and (c) show the inference accuracy for the MNIST and sEMG dataset as a function of the number of trees per forest in the DRF. We follow the training procedure in \cite{zhou2019deep} while varying the number of trees. The DRF is trained at full precision, and the branch-split decision boundary is quantized post-training to evaluate the impact of the boundary precision.
For both models, the accuracy saturates when more than eight trees per forest are utilized. An accuracy of 99.2\% is achievable for the MNIST dataset which is on par with a 3-layer Convolutional Deep Belief Network \cite{CDBN}. For sEMG, the accuracy of the deep random forest model is 72\%, significantly outperforming an advanced LSTM machine learning model \cite{zhou2019deep}. 
These results demonstrate the competitive performance of DRF in performing different classification tasks.

\begin{figurehere}
 \vspace{-2ex}
 \begin{center}
  \includegraphics[width=1\textwidth]{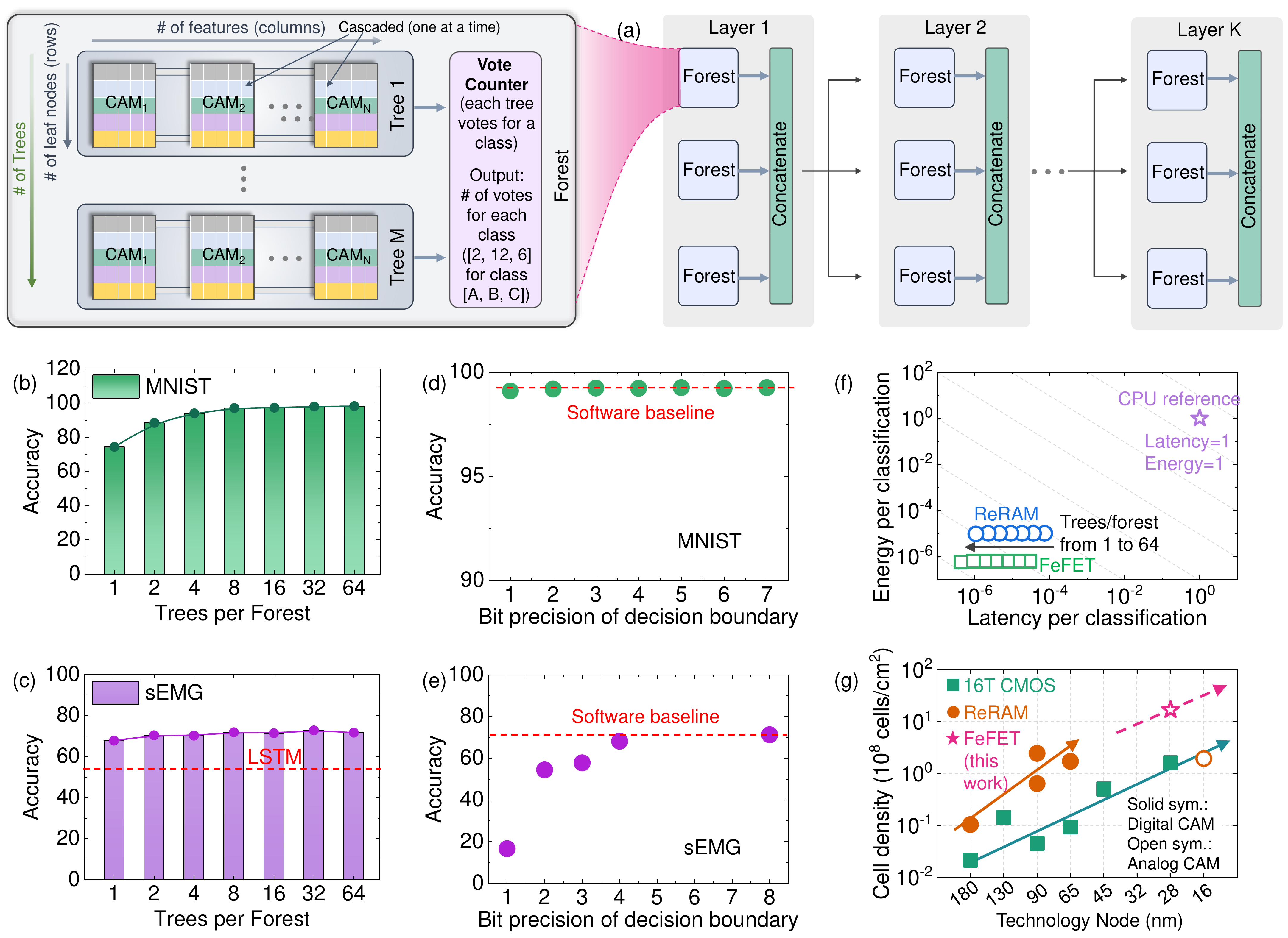}
  \caption{\textit{Benchmarking of the DRF using ferroelectric ACAM arrays. (a) Mapping of the DRF onto ferroelectric ACAM arrays. Each tree of a forest is mapped to an ACAM array where the number of rows corresponds to the number of leaf nodes (i.e., branches) and the number of columns corresponds to the total required features. (b) and (c) Inference accuracy for the MNIST and sEMG dataset with respect to the number of trees per forest, respectively. Excellent accuracy is obtained with the DRF, even when compared with the LSTM models. (d) and (e) Accuracy when mapped to the ACAM array considering the limited precision of the branch-split decision boundary. (f) Energy versus latency for a single classification using the DRF when mapped to the CPU, ReRAM- and FeFET-based ACAMs, respectively. FeFET based ACAM shows superior performance. (g) ACAM cell density, including both the digital and analog cells. The 2FeFET based ACAM achieves the highest density due to its compactness.}}
  \label{fig:system_benchmarking}
 \end{center}
\end{figurehere}

As FeFET ACAM cell can currently hold three bits of \textit{V}\textsubscript{TH} states in this work (per Fig. \ref{fig:supple_fefet_states}), the impact of precision on inference accuracy is evaluated.
Fig. \ref{fig:system_benchmarking}(d) and (e) show the inference accuracy as a function of precision of the decision boundary for the MNIST and sEMG dataset, respectively. 
For MNIST, each grayscale pixel intensity is used as a feature, i.e., non-leaf branch-split node. Since relevant features are either black or white, the DRF performs well even at 1-bit precision. 
However, for the sEMG dataset, the accuracy starts to degrade when the decision boundary precision drops below 4 bits and accuracy is especially low at 1-bit precision. 
The FeFET ACAM with 3-bit precision demonstrated in this work suffers accuracy degradation but still performs better than LSTM for the sEMG dataset. 
Note that due to the core tree structure in a DRF, the higher precision branch-split operation can be realized using ACAM cells with lower precision at the cost of additional ACAM area and energy consumption. 
To implement a higher precision deep random forest, each tree node or some critical nodes (i.e., requiring a higher precision) can be split into multiple tree nodes (lower precision), as illustrated in Fig. \ref{fig:precision} as an example. Each feature must be split into its most significant bits (MSB) and least significant bits (LSB) and treated as two separate features and searched separately. This results in an increased number of branches, and hence the number of rows when mapping to ACAM arrays. In future work, we will evaluate the tradeoffs of extending the precision for FeFET ACAMs.

It is also important to evaluate the impact of device-to-device variation of FeFETs on the classification accuracy of the DRF. The variation in FeFET \textit{V}\textsubscript{TH} (per Fig. \ref{fig:supple_fefet_states}), is directly translated into the variation in the decision boundary, which impacts the accuracy of the branch-split operations. As \textit{V}\textsubscript{TH} variation increases, overlap between neighboring decision boundaries is expected. The impact of such variations may vary by datasets. For MNIST, because the input is binary, the DRF is highly robust to  variation as long as the decision boundary between the black and white pixels is well defined. For sEMG, the input values are not binary, but have intermediate values, which increase the susceptibility of the system to FeFET variation. However, as suggested in Fig. \ref{fig:variation}, when the standard deviation of the decision boundary is less than 7\% of the overall memory window, the accuracy remains unaffected. Considering that the current FeFET \textit{V}\textsubscript{TH} standard deviation is on average 4\% of the overall memory window, DRF that leverages even current devices still yields negligible accuracy loss, demonstrating great robustness. As the FeFET technology continues to improve, variations will be further suppressed \cite{beyer2020fefet}, thus FeFETs will become an even more robust technology platform for DRF implementation.   

To compare FeFET ACAM based DRF with alternative DRF implementations, the ferroelectric ACAM array performance extracted from the simulations in Fig. \ref{fig:spicesim} is used for system-level benchmarking. 
We assume an ACAM array of size 128$\times$128
as the basic ACAM module and that multiple ACAM arrays are cascaded to complete all system-level tasks. 
Fig. \ref{fig:system_benchmarking}(f) shows energy versus latency for a single classification. The DRF implementation on an Intel(R) Core(TM) i7-10750H CPU (14 nm node) @ 2.60GHz with 16GB of RAM is used as a reference (i.e., the latency and energy per classification is considered as 1), against which the system implementation using ACAM arrays based on ReRAM (16 nm node \cite{pedretti2021tree}) and FeFETs are benchmarked. Since ReRAM ACAM array has only been proposed for a decision tree implementation and not for DRF \cite{pedretti2021tree}, we take the reported ReRAM ACAM array characteristics and evaluate its performance in implementing the DRF. 
Due to their parallel nature and compact, in-memory computing characteristics, the ferroelectric ACAM array exhibits significant savings in energy and latency when compared with a CPU (e.g., up to $10^6\times$ saving in energy and latency). FeFET based ACAM arrays have lower energy consumption than their ReRAM counterpart due to the elimination of the DC current flowing through the ReRAM ACAM cell. 
These results suggest great promise for the ferroelectric ACAM array when implementing the DRF. In addition, we also implemented a simple random forest model (i.e., no layer-by-layer structure) using the ferroelectric ACAM and evaluated its performance on some EEG \cite{shoeb2009application} and PET \cite{zhang2021deep} dataset. Table S1 summarizes the metrics including cell size, energy and latency per classification using our ferroelectric ACAM based random forest, as well as other advanced machine learning model implementations. Again superior energy-efficiency and latency for a classification operation using the ferroelectric analog CAM array is demonstrated. 

Fig. \ref{fig:system_benchmarking}(g) provides the evolution of CAM cell density as a function of technology nodes. Both the digital and analog CAM cells are included for completeness.
As expected, with technology scaling, CAM cell density continues to improve. Due to its compactness, the ferroelectric ACAM cell (2FeFET) exhibits the highest density so far, greatly outperforming its ReRAM counterpart (6T2R). As a result, the compact ferroelectric ACAM array could well support the acceleration of the DRF model.

\vspace{-4ex}
\section*{Conclusion} \label{sec:conclusion}

In this work, we implemented the DRF with ferroelectric ACAM array by leveraging the  parallelism and in-memory computing capability of the ACAM array. We demonstrated that DRF inference could be efficiently mapped as the associative search operations in ACAM arrays, as the ACAM cell can realize the key branch-split operation of a decision tree in memory by harnessing the analog polarization states within an FeFET. We validated the functionality of the 2FeFET ACAM cell, and the capability of ACAM arrays in identifying the matching region in the high-dimensional search space. Each ACAM row corresponds to a specific branch from the root node to a leaf node in a decision tree. With the proposed ultra-compact ACAM cell, we show that the FeFET ACAM based DRF accelerator exhibits orders of magnitudes improvement in footprint, and inference energy and latency. These results suggest that ferroelectric ACAMs provide a promising hardware platform to implement DRF as an alternative complement to DNNs for achieving edge intelligence with its interpretability, low latency, and superior energy-efficiency.

\bibliography{ref}

\bibliographystyle{Nature}

\section*{Author contributions}

X.Y. and K.N. proposed and supervised the project. C.L. and Q.H. performed the SPICE simulation. F.M., N.L., and T.K. conducted experimental characterization of ACAM cell and array. S.D. and Z.Z. performed single device measurement. M.L. and T.K. conducted the EBSD characterization. A.F.L., Q.Z., W.Y., Z.S., M.N., X.S.H., and C.Z. performed the benchmarking and system evaluation. All authors contributed to write up of the manuscript.

\section*{Competing interests}
The authors declare no competing interests.
\newpage
\renewcommand{\thefigure}{S\arabic{figure}}
\renewcommand{\thetable}{S\arabic{table}}
\setcounter{figure}{0}
\setcounter{table}{0}

\centering
\title{\textbf{\Large Supplementary Materials}}
\begin{flushleft} 
\textbf{Device Fabrication}
\end{flushleft}

\justify
In this paper, the fabricated ferroelectric field effect transistor (FeFET) features a poly-crystalline Si/TiN (2 nm)/doped HfO\textsubscript{2} (8 nm)/SiO\textsubscript{2} (1 nm)/p-Si gate stack. The devices were fabricated using a 28 nm node gate-first high-K metal gate CMOS process on 300 mm silicon wafers. The ferroelectric gate stack process module starts with growth of a thin SiO\textsubscript{2} based interfacial layer, followed by the deposition of an 8 nm thick doped HfO\textsubscript{2}. 
A TiN metal gate electrode was deposited using physical vapor deposition (PVD), on top of which the poly-Si gate electrode is deposited. The source and drain n+ regions were obtained by phosphorous ion implantation, which were then activated by a rapid thermal annealing (RTA) at approximately 1000 $^\circ$C. This step also results in the formation of the ferroelectric orthorhombic phase within the doped HfO\textsubscript{2}.
For all the devices electrically characterized, they all have the same gate length and width dimensions of 1$\mu$m x 1$\mu$m, respectively.

\begin{flushleft} 
\textbf{Electrical Characterization}
\end{flushleft}
The FeFET device characterization was performed with a PXI-Express system from National Instruments, using a PXIe-1095 cassis, NI PXIe-8880 controller, NI PXIe-6570 pin parametric measurement unit (PPMU) and NI PXIe-4143 source measure unit (SMU).
Prior to characterization all FeFETs are preconditioned using the SMUs by cycling them 100 times with the pulses of +4.5 V, -5 V with a pulse length of 500 ns each. Read out of the memory state is done by a step wise increase of the gate voltage in 0.1 V increments while applying 0.1 V to the drain terminal and measuring the current using the PPMU. Bulk and source terminals are tied to ground at all times. The read operation takes approximately 7 ms. The multi-level characterization of individual FeFETs is performed by putting them in a reference state with a gate voltage of -5 V or +4.5 V for 500 ns for erase or program, respectively. After that a single pulse of increasing amplitude is applied for 200 ns. The gate voltage amplitude stepping is set to 100 mV. After each pulse a delay of 2 s is added to ensure sufficient time for charge detrapping after which a readout is performed. This scheme is repeated for the full switching range.
The CAM measurements are performed in an AND-connected array. One CAM cell is constructed by measuring two FeFETs sharing the same connection at their drain terminal, the matchline. Source and bulk terminal are tied to ground at all times. The FeFETs are programmed to the target $V_T$'s individually, applying a single fixed program pulse specific to the target $V_T$. Readout operation is performed similar to the single devices. The ML is kept at 0.1 V while an stepped gate sweep is performed. Using individual PPMU channels the readout is performed on both FeFETs of one CAM cell.

\begin{flushleft} 
\textbf{Transmission-EBSD Characterization}
\end{flushleft}
For transmission-EBSD characterization, also known as transmission Kikuchi diffraction, a 10~nm Si-doped HfO\textsubscript{2} layer was deposited on a silicon wafer with a thin chemical oxide layer. This was carried out using atomic layer deposition with a cycling ratio of 16:1 (Hf:Si). After capping the layer with a 10~nm TiN top electrode, the film was crystallized via rapid thermal annealing at 800°C. A dimpled sample was prepared and analyzed in a scanning electron microscope using a Bruker Optimus TKD detector. An acceleration voltage of 30~kV and a current of 3.2~nA was used.

\newpage
\begin{flushleft} 
\textbf{Multiple \textit{V}\textsubscript{TH} States in FeFET}
\end{flushleft}
\begin{figurehere}
 \begin{center}
  \includegraphics[width=0.7\textwidth]{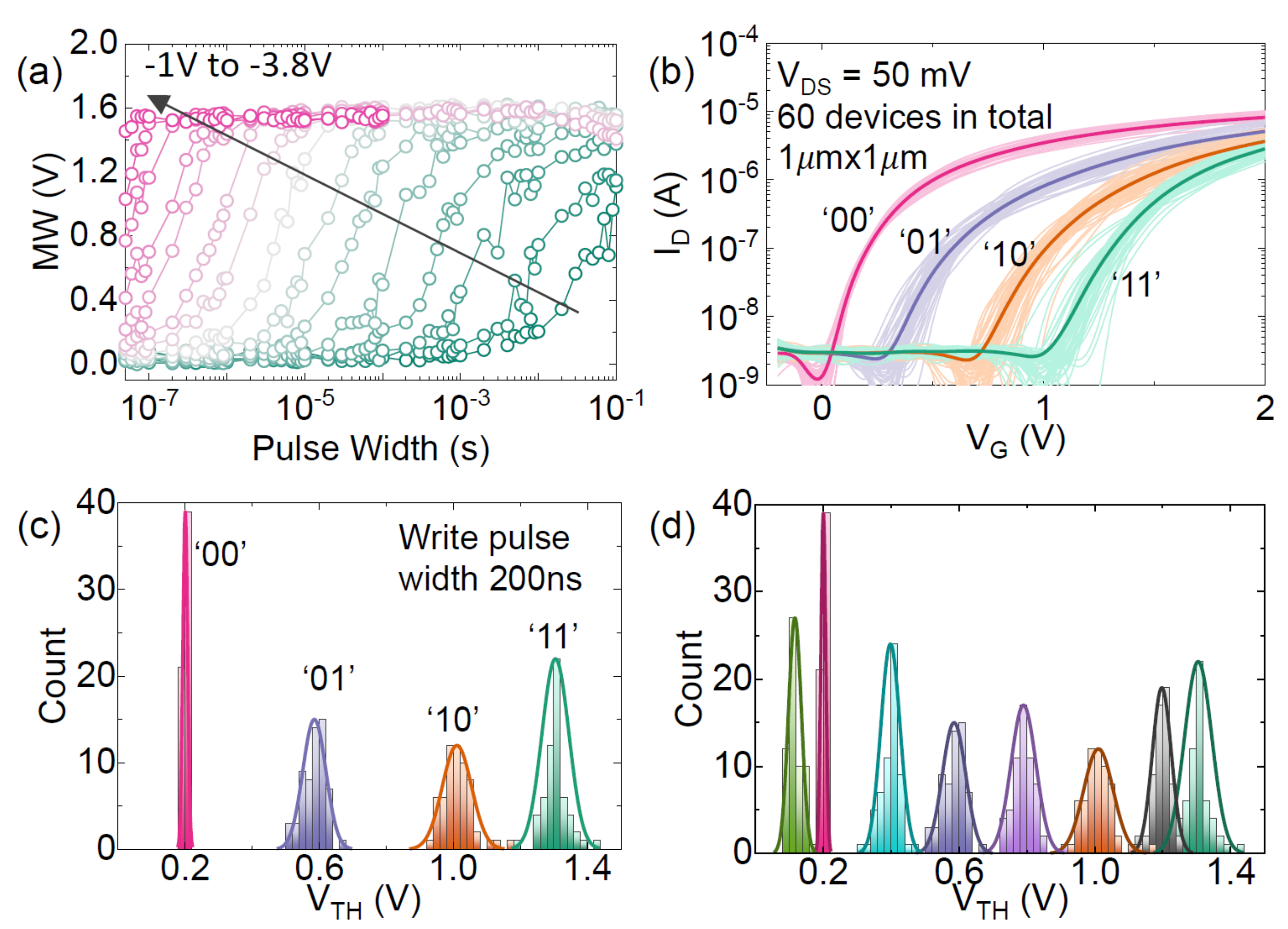}
  \caption{\textit{Multiple states in FeFET. (a) Switching dynamics in FeFET showing the memory window as a function of write pulse width at different write pulse amplitudes. FeFET is initialized with +4V, 1$\mu$s write pulse before each measurement. The write pulse amplitudes changed from -1V to -3.8V with a step of -0.1V. Intermediate \textit{V}\textsubscript{TH} states are observed. (b) \textit{I}\textsubscript{D}-\textit{V}\textsubscript{G} characteristics for four different states in 60 different FeFET devices. (c)/(d) The \textit{V}\textsubscript{TH} distributions for 4/8 levels, respectively. Different levels are reached through a write pulse width different amplitudes. Tight \textit{V}\textsubscript{TH} distribution is obtained given the present unoptimized FeFET devices.}}
  \label{fig:supple_fefet_states}
 \end{center}
\end{figurehere}

\newpage
\begin{flushleft} 
\textbf{Measurement On An 1$\times$16 CAM Array: Other Cells In State 4}
\end{flushleft}
\vspace{-6ex}

\begin{figurehere}
 \begin{center}
  \includegraphics[width=1\textwidth]{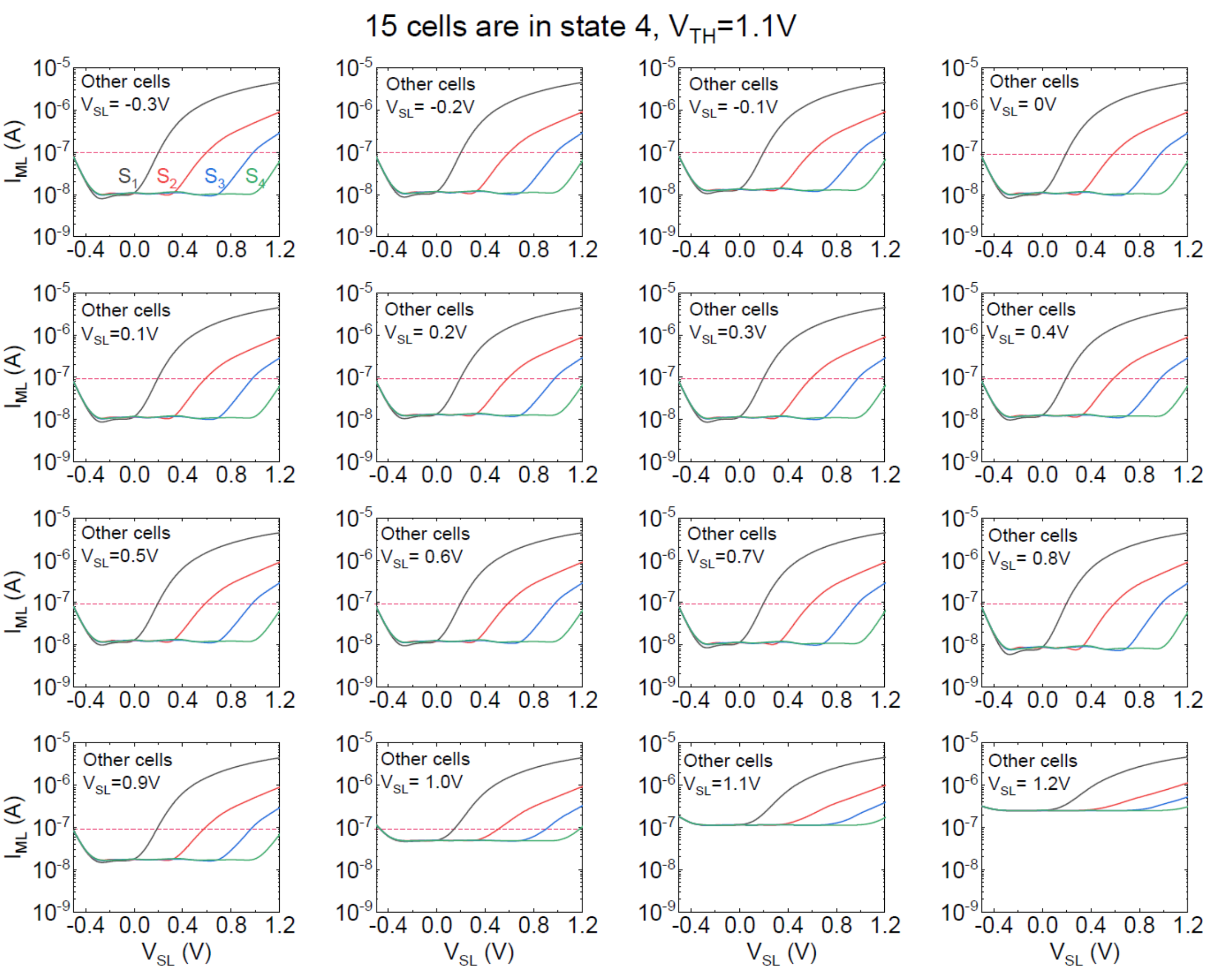}
  \caption{\textit{Measurement on a 1$\times$16 analog CAM array. During testing, all \textit{F}\textsubscript{1} transistors in the array are set to be high-\textit{V}\textsubscript{TH} states, fixing the lower bounds. \textit{F}\textsubscript{0} transistors in 15 cells are configured into the state \textit{S}\textsubscript{4}, i.e., \textit{V}\textsubscript{TH}=1.1V, and the rest target cell is configured to the four \textit{V}\textsubscript{TH} states (black, red, blue, and green curves in each figure). Given this configuration, the target cell \textit{V}\textsubscript{SL} is swept when the other cells are searched with 16 different \textit{V}\textsubscript{SL} values from -0.3V to 1.2V in the step of 0.1V. Assume the current threshold is at $10^{-7}$A (red dashed line), then the matching range can be successfully realized when the other cells are searched with \textit{V}\textsubscript{SL} below \textit{V}\textsubscript{TH} (i.e., $\leq$1.0V) and can be varied on the target cell \textit{V}\textsubscript{SL} dimension.}}
  \label{fig:cellarray_1x16_s4}
 \end{center}
\end{figurehere}

\newpage
\begin{flushleft} 
\textbf{Measurement On An 1$\times$16 CAM Array: Other Cells In State 3}
\end{flushleft}
\vspace{-6ex}

\begin{figurehere}
 \begin{center}
  \includegraphics[width=1\textwidth]{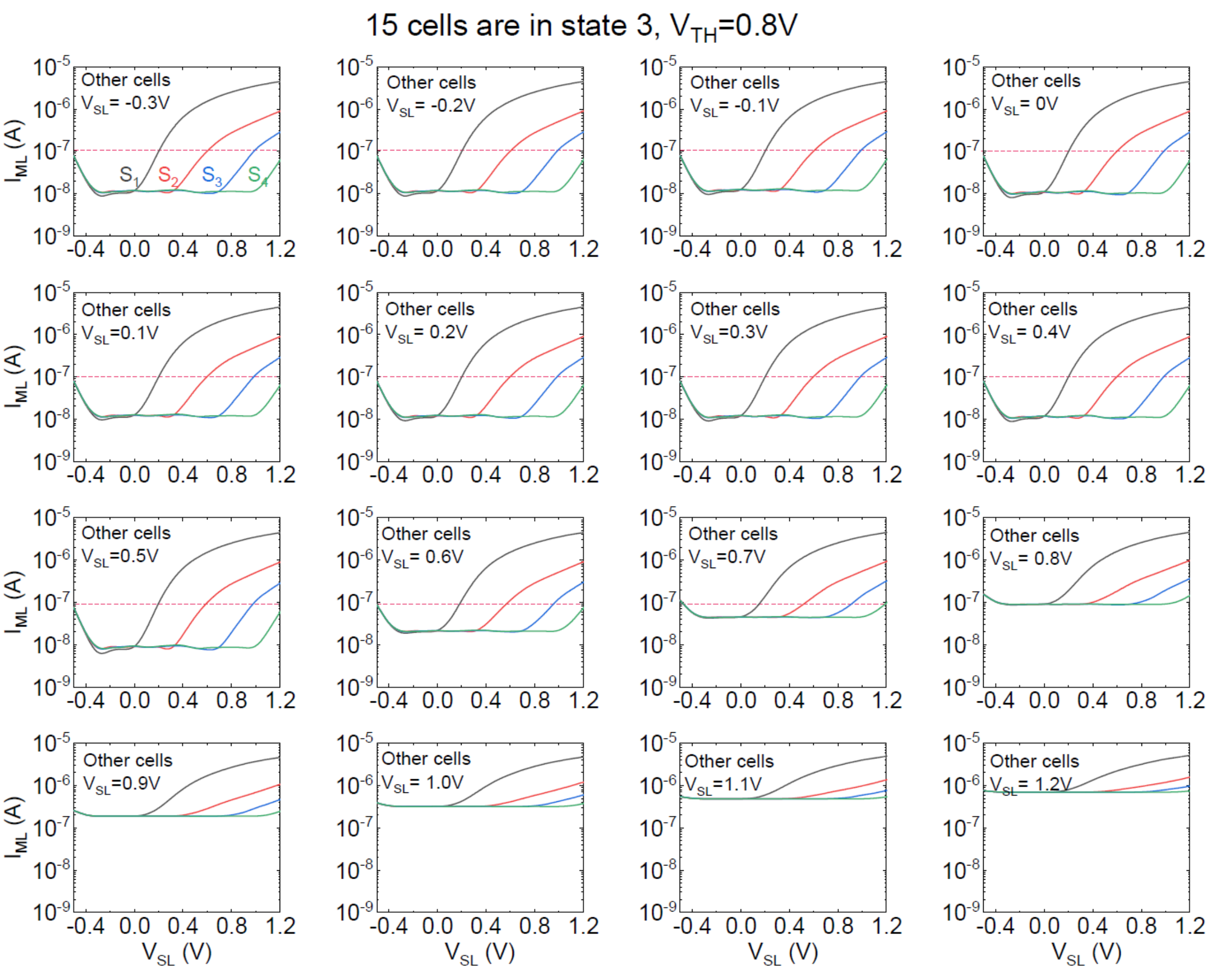}
  \caption{\textit{Measurement on a 1$\times$16 analog CAM array. Similar to Fig.\ref{fig:cellarray_1x16_s4}, all the \textit{F}\textsubscript{1} transistors in the array are set to  high-\textit{V}\textsubscript{TH} states. \textit{F}\textsubscript{0} transistors in 15 cells are configured into the state \textit{S}\textsubscript{3}, i.e., \textit{V}\textsubscript{TH}=0.8V, and the target cell is configured to the four \textit{V}\textsubscript{TH} states (black, red, blue, and green curves in each figure). After the cell configuration, the target cell \textit{V}\textsubscript{SL} is swept when the other cells are searched with 16 different \textit{V}\textsubscript{SL} values from -0.3V to 1.2V in the step of 0.1V. With the current threshold at $10^{-7}$A (red dashed line), the matching range can be now reduced accordingly to \textit{V}\textsubscript{SL}$\leq$0.7V for other cells, and can be varied on the target cell \textit{V}\textsubscript{SL} dimension.}}
  \label{fig:cellarray_1x16_s3}
 \end{center}
\end{figurehere}

\newpage
\begin{flushleft} 
\textbf{Measurement On An 1$\times$16 CAM Array: Other Cells In State 2}
\end{flushleft}
\vspace{-6ex}

\begin{figurehere}
 \begin{center}
  \includegraphics[width=1\textwidth]{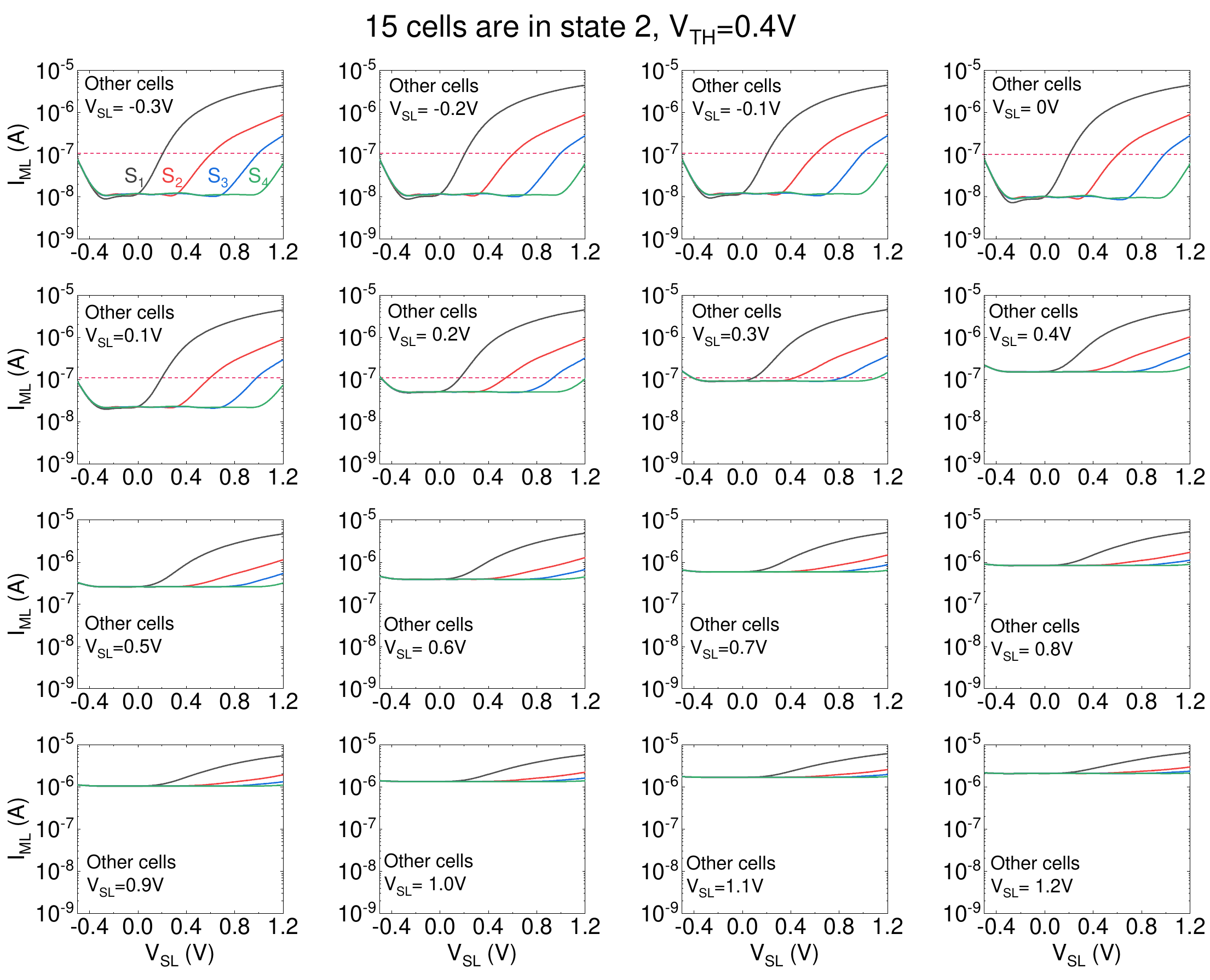}
  \caption{\textit{Measurement on a 1$\times$16 analog CAM array. Similar to Fig.\ref{fig:cellarray_1x16_s4}, all the \textit{F}\textsubscript{1} transistors in the array are set to  high-\textit{V}\textsubscript{TH} states. \textit{F}\textsubscript{0} transistors in 15 cells are configured into the state \textit{S}\textsubscript{2}, i.e., \textit{V}\textsubscript{TH}=0.4V, and the target cell is configured to the four \textit{V}\textsubscript{TH} states (black, red, blue, and green curves in each figure). After the cell configuration, the target cell \textit{V}\textsubscript{SL} is swept when the other cells are searched with 16 different \textit{V}\textsubscript{SL} values from -0.3V to 1.2V in steps of 0.1V. With the current threshold set at $10^{-7}$A (red dashed line), the matching range can now be reduced accordingly to \textit{V}\textsubscript{SL}$\leq$0.3V for other cells, and can be varied on the target cell \textit{V}\textsubscript{SL} dimension.}}
  \label{fig:cellarray_1x16_s2}
 \end{center}
\end{figurehere}

\newpage
\begin{flushleft} 
\textbf{Measurement On An 1$\times$16 CAM Array: Other Cells In State 1}
\end{flushleft}
\vspace{-6ex}

\begin{figurehere}
 \begin{center}
  \includegraphics[width=1\textwidth]{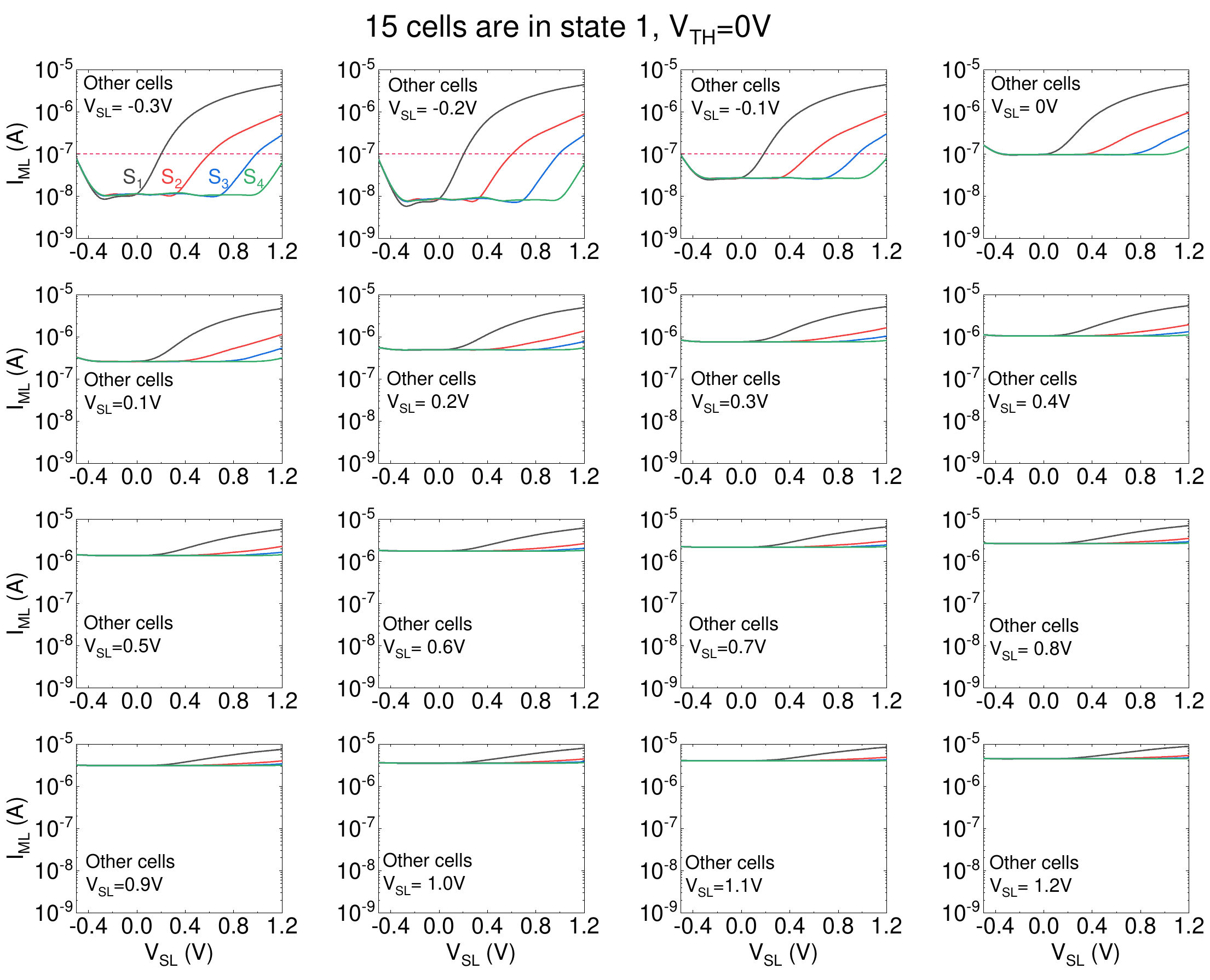}
  \caption{\textit{Measurement on a 1$\times$16 analog CAM array. Similar to Fig.\ref{fig:cellarray_1x16_s4}, all the \textit{F}\textsubscript{1} transistors in the array are set to be high-\textit{V}\textsubscript{TH} states. \textit{F}\textsubscript{0} transistors in 15 cells are configured into the state \textit{S}\textsubscript{3}, i.e., \textit{V}\textsubscript{TH}=0V, and the target cell is configured to the four \textit{V}\textsubscript{TH} states (black, red, blue, and green curves in each figure). After the cell configuration, the target cell \textit{V}\textsubscript{SL} is swept when the other cells are searched with 16 different \textit{V}\textsubscript{SL} values from -0.3V to 1.2V in the step of 0.1V. With the current threshold set at $10^{-7}$A (red dashed line), the matching range can now be reduced accordingly to \textit{V}\textsubscript{SL}$\leq$-0.1V for other cells, and can be varied on the target cell \textit{V}\textsubscript{SL} dimension.}}
  \label{fig:cellarray_1x16_s1}
 \end{center}
\end{figurehere}

\newpage
\begin{flushleft} 
\textbf{SPICE Simulation Setup of Voltage Domain Sensing}
\end{flushleft}
\vspace{-4ex}

\begin{figurehere}
 \begin{center}
  \includegraphics[width=1\textwidth]{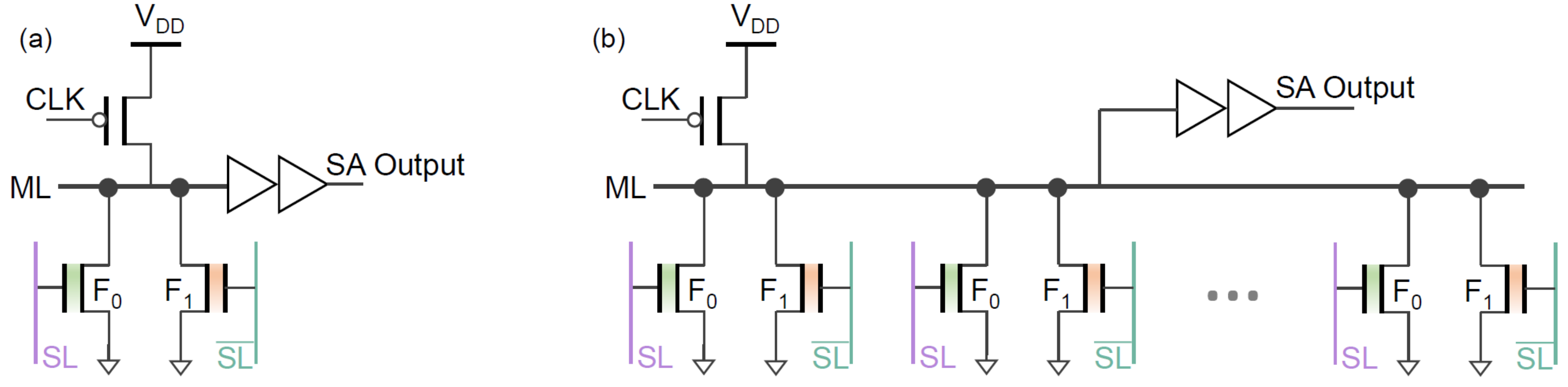}
  \caption{\textit{SPICE simulation setup for voltage domain sensing. (a) Single CAM cell and (b) a CAM word with multiple cells connected on the same match line. A two-stage buffer circuit is adopted as the sense amplifier. The pMOSFET transistor is used to pre-charge the match line for the search operation.} }
  \label{fig:sim_system_setup}
 \end{center}
\end{figurehere}

\newpage
\begin{flushleft} 
\textbf{SPICE Simulation of A Single CAM Cell}
\end{flushleft}
\vspace{-4ex}

\begin{figurehere}
 \begin{center}
  \includegraphics[width=1\textwidth]{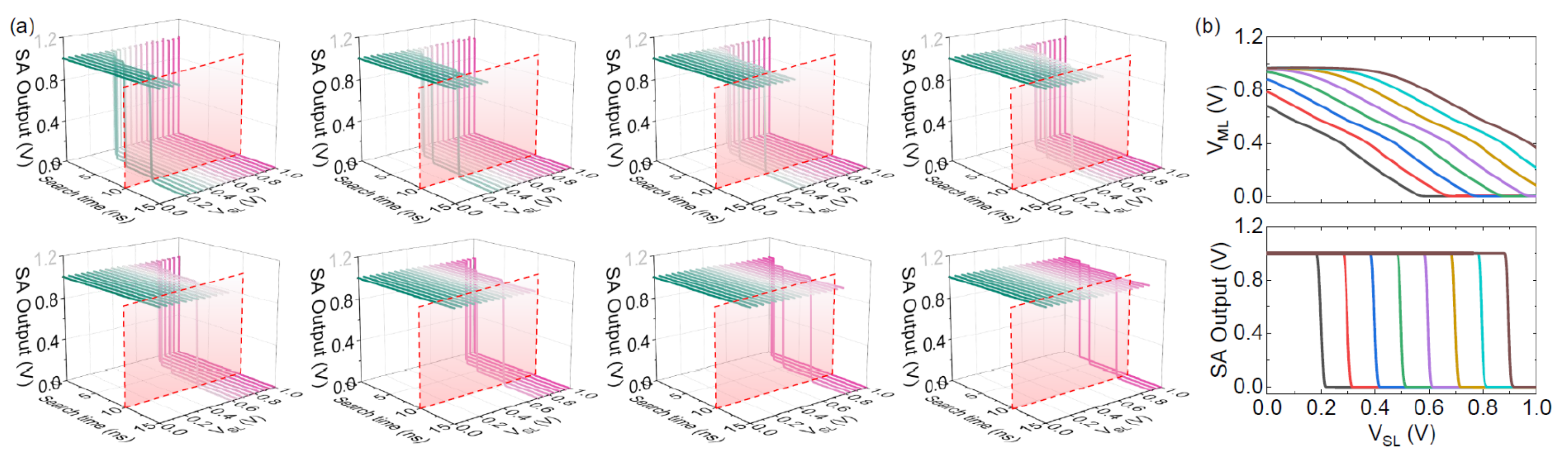}
  \caption{\textit{Transient waveform of the sense amplifier output  in a single CAM cell. (a) Output voltage waveform at different \textit{V}\textsubscript{SL} values for 8 different cell configurations. By configuring the \textit{F}\textsubscript{0} FeFET into 1 of the 8 \textit{V}\textsubscript{TH} states, the decision boundary threshold is shifted accordingly. In this work, a 10ns search time, i.e., the red dashed plane, is chosen to sense the output voltage vs. the \textit{V}\textsubscript{SL}. (b) The match line voltage and the final output voltage as a function of the \textit{V}\textsubscript{SL} sensed at the search time of 10ns. A sharp \textit{V}\textsubscript{SL} decision boundary can be realized with the sensing circuit. }}
  \label{fig:single_cell_sense}
 \end{center}
\end{figurehere}

\newpage
\begin{flushleft} 
\textbf{SPICE Simulation of Current Sensing in An 1$\times$2 CAM Array}
\end{flushleft}
\vspace{-6ex}

\begin{figurehere}
 \begin{center}
  \includegraphics[width=1\textwidth]{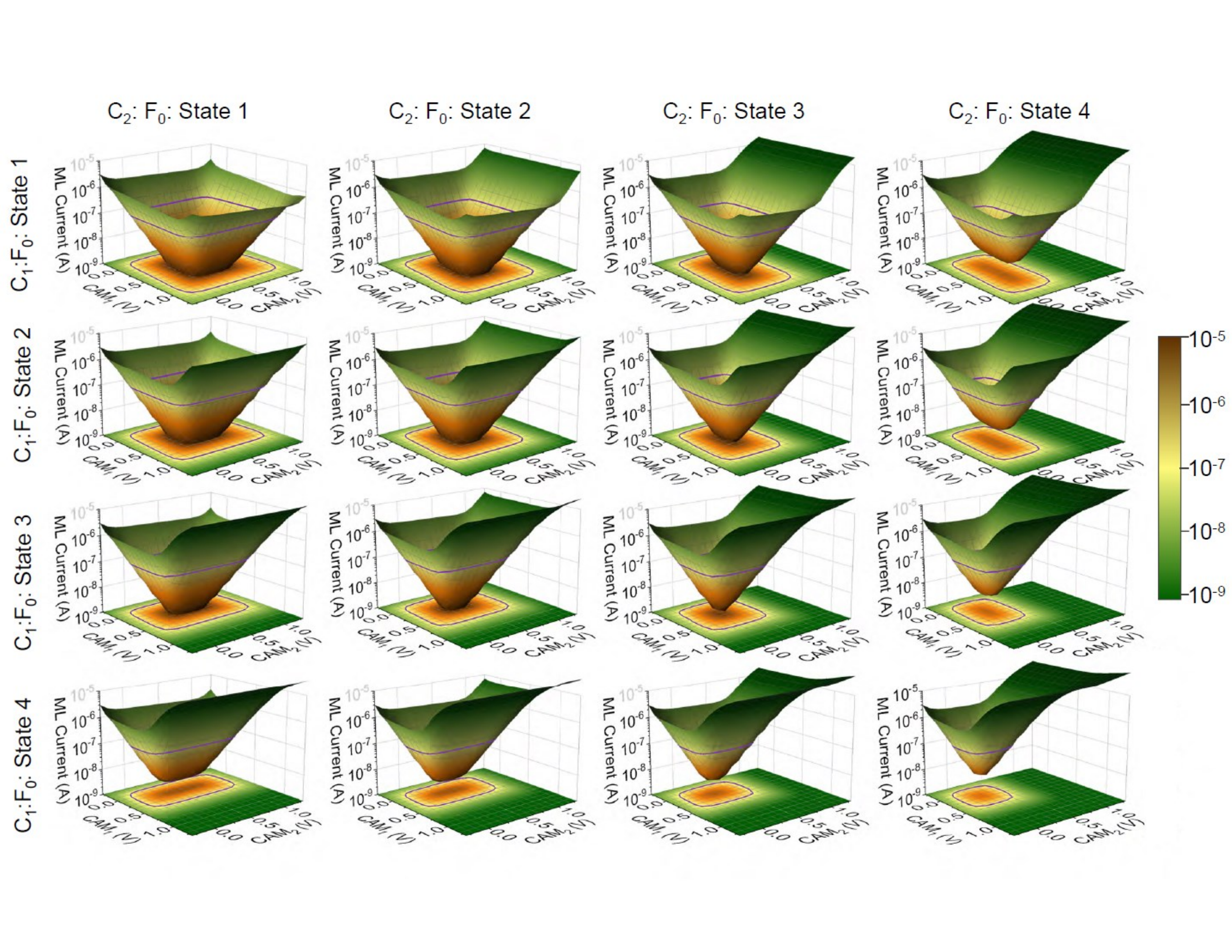}
  \vspace{-8ex}
  \caption{\textit{Current sensing in a 1$\times$2 CAM array. Similar to the experimental measurement on with 1$\times$2 CAM array shown in Fig.\ref{fig:singlecell_exp}, a 1$\times$2 CAM array is simulated in SPICE and the match line current is measured when the array is configured into 1 of the 4$\times$4 configurations. Similar to the experiment, the simulation confirms that the decision boundary on each \textit{V}\textsubscript{SL} dimension is independent of each other and together the \textit{V}\textsubscript{SL}'s of all the cells define the matching region in the \textit{V}\textsubscript{SL} space.}}
  \label{fig:sim_1x2_array_current}
 \end{center}
\end{figurehere}

\newpage
\begin{flushleft} 
\textbf{SPICE Simulation of Voltage Domain Sensing in An CAM Array}
\end{flushleft}
\vspace{-6ex}

\begin{figurehere}
 \begin{center}
  \includegraphics[width=1\textwidth]{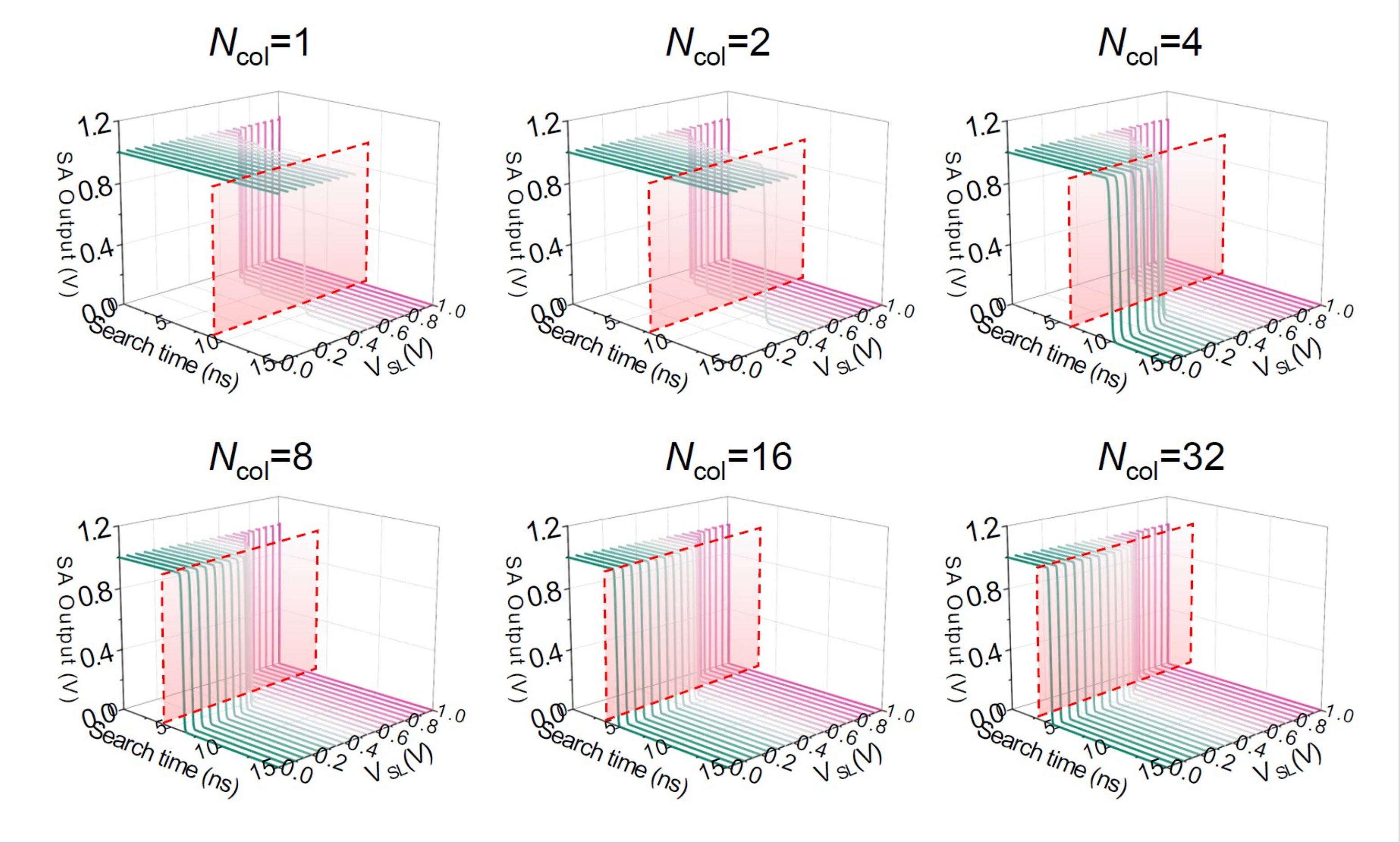}
  \caption{\textit{Voltage domain sensing of a CAM array with different numbers of columns. In the simulation, the worst sensing scenario is considered where only one cell is swept, its decision boundary is set to be close to 0.5V,  all other cells connected on the same match line store the same state, and all other cells are searched with the same \textit{V}\textsubscript{SL} (close to their decision boundary). With more cells connected to the match line, the leakage current becomes larger as contributed by  other cells, thus advancing the discharge of the match line and the sense amplifier output. Therefore the search time, as indicated by the red dashed plane, needs to be adjusted for a CAM array with a larger number of columns.}}
  \label{fig:sim_array_col}
 \end{center}
\end{figurehere}

\newpage
\begin{flushleft} 
\textbf{Branch Split Threshold Precision Extension}
\end{flushleft}
\vspace{-4ex}

\begin{figurehere}
 \begin{center}
  \includegraphics[width=0.9\textwidth]{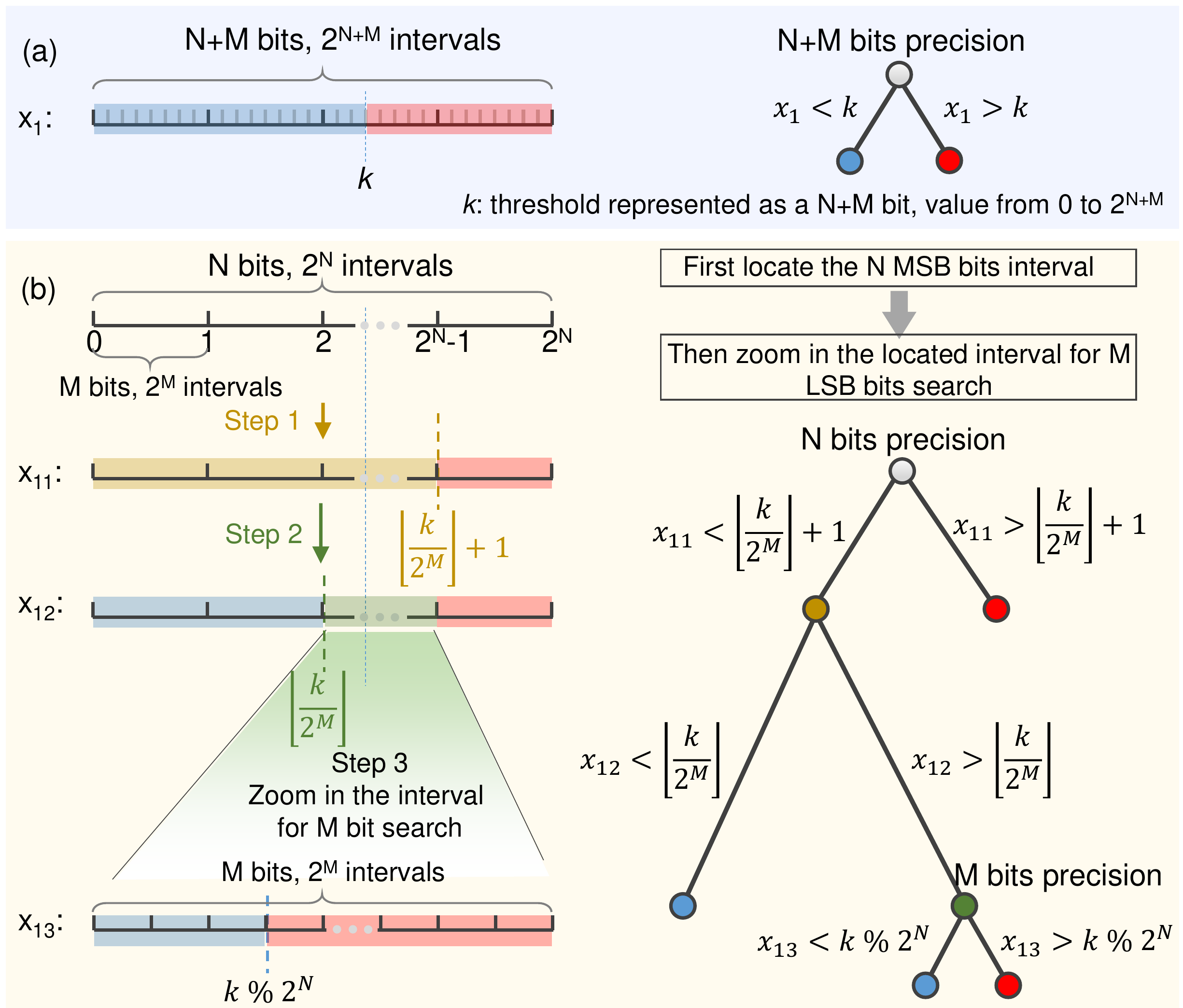}
  \caption{\textit{A possible approach to extend the precision of the branch split threshold using ferroelectric analog CAM cells with a limited precision. (a) The branch splits based on the threshold value $k$, which is represented as $N+M$ bits. The blue/red are the less-than/greater-than branches, respectively. (b) Realizing the $N+M$ bit split threshold precision utilizing only analog CAM cells with $N$ and $M$ bits precision. The key idea is to first locate the interval identified by $N$ MSB bits (the green interval), then the identified interval is used for $M$ LSB bits search. To identify the $N$ bit interval, two steps are used, where step 1 is to find the upper bound of the interval and step 2 is to find the lower bound of the interval. Then step 3 is to zoom in the interval for $M$ bits search.}}
  \label{fig:precision}
 \end{center}
\end{figurehere}


\newpage
\begin{flushleft} 
\textbf{Impact of Device-to-Device Variation on The System Accuracy}
\end{flushleft}
\vspace{-4ex}

\begin{figurehere}
 \begin{center}
  \includegraphics[width=0.9\textwidth]{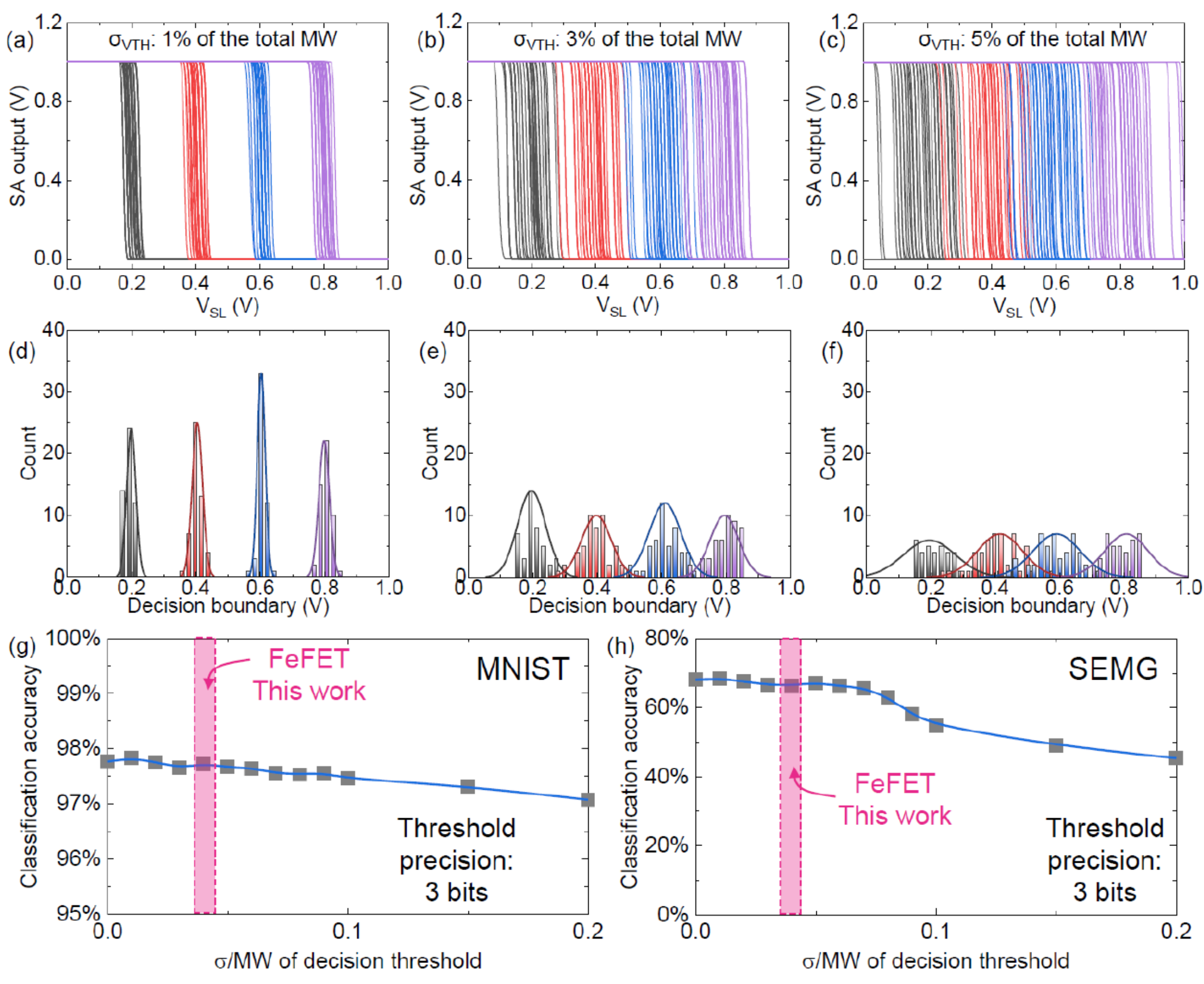}
  \caption{\textit{Impact of device-to-device variation on the accuracy of deep random forest. (a), (b), (c) are the simulated sense amplifier output of a single CAM cell given a FeFET \textit{V}\textsubscript{TH} standard deviation of 1\%, 3\%, and 5\% of the overall memory window, respectively. (d), (e), (f) are the histograms of the decision boundaries under 1\%, 3\%, and 5\% variation. The variation in the decision boundary degrades as \textit{V}\textsubscript{TH} variation increases. Studies of the deep random forest accuracy under different degrees of variation in the decision boundary for (g) the MNIST and (h) the SEMG dataset. For MNIST, the input is binary, which is highly robust to the decision boundary variation as long as correct distinction between the black and white pixels can be made. For the SEMG, accuracy degradation starts to emerge when the standard deviation of the decision boundary exceeds 7\% of the overall memory window. FeFET this work exhibits a standard deviation of 4\% of the overall memory window, as shown in Fig.\ref{fig:supple_fefet_states}.}}
  \label{fig:variation}
 \end{center}
\end{figurehere}

\newpage
\begin{flushleft} 
\textbf{Compilation of Advanced Machine Learning Model Hardware}
\end{flushleft}

In addition to the DRF discussed in the main text, we also implemented the random forest (i.e., no layer-by-layer structure and only a single forest) using the ferroelectric analog CAM. 
The performance of the implementation is evaluated with seizure detection on the EEG (CHB-MIT) \cite{ref14} and PET/CT \cite{ref15} datasets for pediatric patients.
3-bit precision is used to quantify the extracted features for the training of the random forest model. A 128$\times$128 array is used to implement the trained RF models, accommodating all the features of PET/CT and EEG datasets. A compilation of various reported hardware for machine learning models are presented here. The intention is not to compare against different hardware implementations as direct comparison is unfair due to different models and technologies used.

\newcolumntype{C}[1]{>{\centering}m{#1}}

\begin{table}[h]
    \centering
{
\fontfamily{phv}\selectfont
\tiny
\rowcolors{1}{black!5}{black!5}
\extrarowheight=\aboverulesep
\addtolength{\extrarowheight}{\belowrulesep}
\aboverulesep=0pt
\belowrulesep=0pt
\begin{tabular}{C{1cm} C{2cm} C{2cm} C{1.4cm} C{1cm} C{2cm} C{1.5cm} C{1.5cm} C{0cm}}

\toprule
\textbf{Reference} & \textbf{Machine Learning Model} & \textbf{Architecture} & \textbf{Technology} & \textbf{Bit cell size} & \textbf{Dataset} & \textbf{Energy} / \textbf{Classification} & \textbf{Classification time} & \\

\midrule


\cite{ref1} & RF & Intel X5560 & CMOS 45nm & N/A & URL Reputation \cite{ref9} & 20.4mJ & 107.5$\mu$s & \\ \arrayrulecolor{black!10} \midrule
\cite{ref1} & RF & NVIDIA Tesla M2050 & CMOS 45nm & N/A & URL Reputation & 11mJ & 49$\mu$s & \\ \arrayrulecolor{black!10} \midrule
\cite{ref1} & RF & Xilinx Virtex-6 & CMOS 40nm & N/A & URL Reputation  & 0.351mJ & 31.9$\mu$s & \\ \arrayrulecolor{black!10} \midrule

\cite{ref2} & Vocabulary tree & Digital  & CMOS 65nm & N/A & COIL-100 \cite{ref10} & 460$\mu$J & 16.7ms & \\ \arrayrulecolor{black!10} \midrule

\cite{ref3} & Vocabulary tree & Digital & CMOS 65nm & N/A & N/A & 186.7$\mu$J & 33.3ms & \\ \arrayrulecolor{black!10} \midrule
\cite{ref4} & AdaBoost & In-memory & CMOS 180nm &4.33$\mu$m$^2$ & MNIST \cite{ref11} & 0.6nJ & 20ns & \\ \arrayrulecolor{black!10} \midrule

\cite{ref5} & SVM & In-memory & CMOS 65nm & 1.94$\mu$m$^2$ & MIT CBCL \cite{ref12} & 963pJ & 107.5ns & \\ \arrayrulecolor{black!10} \midrule

\cite{ref6} & SVM & In-memory & CMOS 65nm & 2.56$\mu$m$^2$ & MIT CBCL& 42pJ & 31.2ns & \\ \arrayrulecolor{black!10} \midrule

\cite{ref7} & RF & In-memory & CMOS 65nm & 1.94$\mu$m$^2$ & MIT CBCL & 19.4nJ & 2.7$\mu$s & \\
\arrayrulecolor{black!10} \midrule

\cite{ref8} & RF & In-Memory & RRAM + CMOS 16nm & 0.52$\mu$m$^2$ & IRIS \cite{ref13} & 0.17nJ & 48ns & \\ \arrayrulecolor{black!10} \midrule

\textbf{This work} & RF & In-Memory & FeFET + CMOS 28nm & 0.06$\mu$m$^2$ & EEG \cite{ref14} / PET \cite{ref15}& 2.91pJ & 1.9ns & \\ 

\arrayrulecolor{black}\bottomrule

\end{tabular}
}


\caption{Performance summary of advanced machine learning model implementations. The random forest hardware based on ferroelectric analog CAM is compact and energy-efficient.}
\label{tab:array}
\end{table}

\end{document}